\documentclass[journal=jctcce,
manuscript=article,
layout=onecolumn]{achemso}
\setkeys{acs}{articletitle = true}
\SectionNumbersOn

\DeclareUnicodeCharacter{2009}{\,}

\usepackage{graphicx}
\usepackage{epsfig}
\usepackage{amsmath}
\usepackage{amssymb}
\usepackage{graphics}
\usepackage{latexsym}
\usepackage{subfigure}
\usepackage{xcolor}
\usepackage[update,prepend]{epstopdf}
\usepackage{booktabs}
\usepackage{multirow}

\title{NMR relaxation rates of quadrupolar aqueous ions from classical molecular dynamics using force-field specific Sternheimer factors}

\author{Iurii Chubak}
\affiliation{Sorbonne Universit{\'e} CNRS, Physico-Chimie des {\'e}lectrolytes et Nanosyst{\`e}mes Interfaciaux, F-75005 Paris, France}
\author{Laura Scalfi}
\affiliation{Sorbonne Universit{\'e} CNRS, Physico-Chimie des {\'e}lectrolytes et Nanosyst{\`e}mes Interfaciaux, F-75005 Paris, France}

\author{Antoine Carof}
\affiliation{Universite de Lorraine, CNRS, LPCT, F-54000, 
Nancy, France}

\author{Benjamin Rotenberg}
\affiliation{Sorbonne Universit{\'e} CNRS, Physico-Chimie des {\'e}lectrolytes et Nanosyst{\`e}mes Interfaciaux, F-75005 Paris, France}
\email{benjamin.rotenberg@sorbonne-universite.fr}

\begin{document}

\maketitle

\begin{abstract}
The nuclear magnetic resonance (NMR) relaxation of quadrupolar nuclei is governed by the electric field 
gradient (EFG) fluctuations at their position. In classical molecular dynamics (MD), the electron cloud 
contribution to the EFG can be included \emph{via} the Sternheimer approximation, in which the full EFG at 
the nucleus that can be computed using quantum density functional theory (DFT) is considered to be proportional to that arising from the external,
classical charge distribution. In this work, we systematically assess the quality of the Sternheimer 
approximation as well as the impact of the classical force field (FF) on the NMR relaxation rates of 
aqueous quadrupolar ions at infinite dilution. In particular, we compare the rates obtained using an 
ab initio parametrized polarizable FF, a recently developed empirical FF with scaled ionic charges and a simple empirical non-polarizable FF with formal ionic charges. Surprisingly, all three FFs considered 
yield good values for the rates of smaller and less polarizable solutes (Li$^+$, Na$^+$, K$^+$, Cl$^-$), provided 
that a model-specific Sternheimer parametrization is employed. Yet, the polarizable and scaled charge 
FFs yield better estimates for divalent and more polarizable species (Mg$^{2+}$, Ca$^{2+}$, Cs$^{+}$). 
We find that a linear relationship between the quantum and classical EFGs holds well in all of the 
cases considered, however, such an approximation often leads to quite large errors in the resulting EFG 
variance, which is directly proportional to the computed rate. 
We attempted to reduce the errors by including first order nonlinear corrections to the EFG, 
yet no clear improvement for the resulting variance has been found. The latter result indicates that 
more refined methods for determining the EFG at the ion position, in particular those that take into account the
instantaneous atomic environment around an ion, might be necessary to systematically improve the NMR 
relaxation rate estimates in classical MD. 
\end{abstract} 

\section{Introduction}

In a typical nuclear magnetic resonance (NMR) experiment,
the nuclei with spin $I \geq 1$ predominantly relax through 
the quadrupolar mechanism that couples the quadrupolar moment
of the nucleus $eQ$ with the electric field gradient (EFG) $V_{\alpha \beta}$ 
at its position \cite{abragam1961principles}. Thus, 
the description of quadrupolar NMR relaxation is pertinent
to the majority of alkaline
metal, alkaline earth metal, and halide ions, that are
species of considerable chemical, technological, and biological
interest ($^7{\rm Li}^+$, $^{23}{\rm Na}^+$, $^{39}{\rm K}^+$,
$^{25}{\rm Mg}^{2+}$, $^{35}{\rm Cl}^-$, etc). 
During an NMR experiment, nuclear magnetic moments precess with the characteristic Larmor 
frequency $\omega_0 = \gamma B_0$ where $\gamma$ is the gyromagnetic ratio of a given nucleus and 
$B_0$ is the external magnetic field.
Provided that the typical molecular time scale $\tau$ over which the EFG decorrelates is small compared to $1/\omega_0$
($\omega_0 \tau \ll 1$, the so-called extreme narrowing regime),
the magnetization along both the longitudinal and transverse directions of $B_0$
decays exponentially with equal rate constants $1/T_1$ and $1/T_2$, the spin-lattice
and spin-spin relaxation rates, respectively.
The latter quantities can be then computed as follows\cite{abragam1961principles,RS1993}:
\begin{equation}
\label{eq:nmr_rates}
\frac{1}{T_1} = \frac{1}{T_2} = \frac{1}{20}
\frac{2I+3}{I^2(2I-1)} \left(\frac{eQ}{\hbar}\right)^2 
\int_0^{\infty} {\rm d}t \,
\langle {\mathbf{V}}(t)\,{:}{\mathbf{V}}(0) \rangle,
\end{equation}
where $\hbar$ is the reduced Planck constant, 
${\mathbf{V}}$ is the total EFG tensor at the nucleus position,
and  $\langle {\mathbf{V}}(t)\,{:}{\mathbf{V}}(0) \rangle \equiv
\langle \sum_{\alpha,\beta} V_{\alpha\beta}(t) V_{\alpha\beta}(0) \rangle$
($\alpha$ and $\beta$ run over the three Cartesian components)
is the EFG autocorrelation function (ACF), $C_{\rm EFG}(t)$.
Finally, $\langle \dots \rangle$ in Eq.~\eqref{eq:nmr_rates} 
denotes an average in the canonical ensemble (fixed number of particles $N$, volume $V$ and temperature $T$), without an imposed magnetic field. 
The isotropic character of the equilibrium system permits 
to have such simple formulation of Eq.~\eqref{eq:nmr_rates} that is rooted in the 
linear response theory.

The estimation of the EFG ACF requires both a good sampling of the microscopic dynamics of the system and an accurate computation of the EFG along the trajectory. In principle, ab initio molecular dynamics (MD) can provide both features\cite{Philips2017,Philips2018,Philips2020}, yet the entailed computational cost often limits the ability to converge the corresponding ACF with sufficient accuracy for a reliable estimate of its integral\cite{Badu2013}. This motivated the use of classical MD to sample configurations over the relevant time scales, yet the accurate estimation of the EFG at the nucleus position has 
remained a formidable challenge in classical MD\cite{EngJon1981, EngJon1982, EngJon1984, RS1993, Carof2014, Mohammadi2020}. 
Such challenge comes from the fact that the dominant contribution
to the total EFG $V_{\alpha\beta}$ is due to the intraatomic 
electronic charge distribution around a nucleus that is not available in atomistic classical models and is even difficult to obtain accurately with electronic structure calculations. 
Yet, quantum density functional theory (DFT) allows to quite accurately compute EFGs for 
main group elements and some transition metals\cite{Autschbach2010}, although a complete understanding of the solvation effects on the EFGs has remained challenging. 
In the case of monoatomic ions, the widespread electrostatic 
models of the EFG relaxation\cite{Hertz1973_1,Hertz1973_2,Valiev1961,Friedman} consider that
the EFG at the nucleus is created by the inhomogeneous distribution of external charges that
polarize the electronic cloud of the given ion. 
Consequently, a linear relation between the total $V_{\alpha\beta}$
and external $V^{\rm ext}_{\alpha\beta}$ EFG is often assumed,
yielding the so-called Sternheimer approximation\cite{Sternheimer1950}:	
\begin{equation}
\label{eq:st_approx_eq}
\mathbf{V} \simeq (1 + \gamma)
\mathbf{V}_{\rm ext},
\end{equation}
where the constant $\gamma$ is the Sternheimer (anti-)shielding 
factor \cite{Foley1954, Sternheimer1966} that is usually large, $\gamma \gg 1$
(note that Eq.~\eqref{eq:st_approx_eq} uses a different sign convention for $\gamma$
than in Refs.~\cite{Foley1954, Sternheimer1966, Schmidt1980}). 
While the exact external charge distribution that comes from solvent molecules and
gives rise to $\mathbf{V}_{\rm ext}$ is generally unknown, in classical MD it is approximated by a simpler distribution corresponding to the charges used to compute electrostatic interactions. Following previous work, we will assume
that $V_{\alpha\beta}^{\rm ext}$ in Eq.~\eqref{eq:st_approx_eq} is due to the external, force-field based charge distribution readily available in classical MD,
whereas a specific value of $\gamma$  can be determined using auxiliary quantum-mechanical calculations 
for the system at hand. For instance, the Sternheimer factors can be
derived using a perturbation theory in the long-range approximation\cite{Foley1954, Sternheimer1966} 
(usually denoted as $\gamma_{\infty}$), in which the polarization
of a spherical electron cloud around an ion is due to a distant 
charge distribution. Nevertheless, the values of $\gamma_{\infty}$
that do not capture polarization effects from solvent-solute
interactions were shown to differ considerably from that of ions in aqueous solutions \cite{Carof2014}. 
Inserting Eq.~\eqref{eq:st_approx_eq} into Eq.~\eqref{eq:nmr_rates},
the expression \eqref{eq:nmr_rates} for the relaxation rate can be recast as
\begin{equation}
\label{eq:nmr_rates2}
\frac{1}{T_1} = C_Q (1+\gamma)^2 \langle \mathbf{V}_{\rm ext}^2 \rangle \tau_c,
\end{equation}
where $C_Q \equiv \frac{1}{20}
\frac{2I+3}{I^2(2I-1)} \left(\frac{eQ}{\hbar}\right)^2$
is an ion specific constant, 
$\langle \mathbf{V}_{\rm ext}^2 \rangle \equiv 
\langle {\mathbf{V}}_{\rm ext}(0)\,{:}{\mathbf{V}}_{\rm ext}(0) \rangle$ is the variance of the external EFG, and
$\tau_c$ is an effective correlation time
\begin{equation}
\label{eq:tauc}
\tau_c = \langle \mathbf{V}_{\rm ext}^2 \rangle^{-1} \,
\int_0^{\infty} {\rm d}t \,
\langle {\mathbf{V}}_{\rm ext}(t)\,{:}{\mathbf{V}}_{\rm ext}(0) \rangle
\; .
\end{equation}
Equations \eqref{eq:nmr_rates2} and \eqref{eq:tauc}
serve as a starting point of the present work. Using a series of classical
force fields (FFs) as described in detail below in Section~\ref{sec:sim_details}, 
we simulate single ions dissolved in 
water to determine the variance of the external EFG 
$\langle \mathbf{V}_{\rm ext}^2 \rangle$, the ACF of external EFG, 
and the effective correlation time $\tau_c$ \eqref{eq:tauc}. 
In line with the earlier work \cite{Carof2014}, we then determine
the effective model-specific Sternheimer factors $\gamma_{\rm eff}$ by comparing the EFG at the ion position obtained from classical
MD to that from ab initio (AI) DFT calculations on a set of configurations generated with classical FFs. Finally, the resulting combination
of $\langle \mathbf{V}_{\rm ext}^2 \rangle$, $\tau_c$, and $\gamma_{\rm eff}$, and ionic parameters ($I$, $eQ$) allows to consistently determine the NMR 
relaxation rate $1/T_1$ as given by Eq.~\eqref{eq:nmr_rates2}.

In this work we aim to address the following questions:
$(i)$ Can accurate quadrupolar NMR relaxation rates be obtained  from classical MD simulations with various degrees of sophistication, provided that the electronic 
contribution to the EFG is included via 
AI-parametrized, model-specific Sternheimer factors that
capture local solvent-solute interactions?
$(ii)$ How well does the Sternheimer approximation
describe the AI-derived EFGs at the ion position and 
can the incorporation of non-linear effects improve 
the predictions of the Sternheimer model?
$(iii)$ How do the Sternheimer factors depend on the structure
of the hydration shell around an ion?
$(iv)$ To what extent the treatment of polarizability 
in classical MD impacts the obtained NMR relaxation rates?
The rest of the article is therefore structured as follows. In Section~\ref{sec:sim_details}, we describe the details of our classical MD and 
quantum DFT simulations. In Section~\ref{sec:stern_approx},
we validate the Sternheimer approximation for different ions using
a series of classical FFs and consider its possible extensions as well as dependence
on the hydration shell structure around an ion. 
In Section~\ref{sec:nmr_rates}, we consider the EFG relaxation
at the nucleus position of various alkali metal, alkaline earth metal, and halide ions with the employed FFs to determine their quadrupolar NMR
relaxation rates at infinite dilution, before concluding in Section~\ref{sec:disc}.

\section{Simulation details}
\label{sec:sim_details}

To systematically assess the impact of the underlying 
classical model on the resulting NMR relaxation rates,
we investigated three FFs for aqueous electrolyte solutions
that differ in terms of 
their parametrization strategy (AI- vs. empirically-parametrized),
the geometry of water molecules, and the treatment of electronic polarizability:
$(i)$ the explicitly polarizable ion model\cite{PIM} (PIM) based 
on the rigid four site Dang-Chang water\cite{DangChang};
$(ii)$ the Madrid-2019 model\cite{Madrid_2019} with scaled 
ionic charges based on the rigid four site TIP4P/2005 water\cite{TIP4P2005};
$(iii)$ the Amber-14 FF \citep{Joung2008,Joung2009,Li2013,AMBER14}
based on the rigid three site SPC/E water\cite{SPCE} 
and formal ionic charges.
In both Dang-Chang and TIP4P/2005 rigid water models, the molecules carry an additional massless M-site that bears the 
negative charge and lies 
along the bisector of the HOH bond angle. In addition, the M-site of the Dang-Chang
water carries an induced dipole moment in the PIM, as discussed
further below. 

The total potential energy of the PIM is given by\cite{PIM}:
\begin{equation}
\label{eq:UPIM}
U^{\rm PIM}_{\rm tot} = U_{\rm charge} + 
U_{\rm disp} + U_{\rm rep} + U_{\rm LJ} + U_{\rm pol}.
\end{equation}
In Eq.~\eqref{eq:UPIM}, $U_{\rm charge}$ stands for the Coulomb
interaction between charges $q^i$ and $q^j$ separated by
the distance $r_{ij}$ (unless otherwise specified, atomic units are employed throughout the article):
\begin{equation}
\label{eq:U_charge}
U_{\rm charge} = \sum_{i < j} \frac{q^i q^j}{r_{ij}}.
\end{equation}
Note that in the PIM ions bear formal charges (+1, -1, +2), while the sites in Dang-Chang water molecules bear partial charges ($q_{\rm H} = +0.5190$, $q_{\rm O} = 0$, $q_{\rm M} = -2q_{\rm H}$).
$U_{\rm disp}$ is the dispersion potential:
\begin{equation}
\label{eq:U_disp}
U_{\rm disp} = - \sum_{i < j} \left[
f^{ij}_6(r_{ij}) \frac{C^{ij}_6}{r_{ij}^6} +
f^{ij}_8(r_{ij}) \frac{C^{ij}_8}{r_{ij}^8}
\right],
\end{equation}
where $f^{ij}_n = 1 - \exp\left({-b^{ij}_D r_{ij}}\right) 
\sum_{k = 0}^{n} \frac{\left(b^{ij}_D r_{ij}\right)^k}{k!}$
are the short-range corrections of the Tang-Toennies type.
$U_{\rm rep}$ captures the repulsive part of ion-ion and 
ion-water interactions:
\begin{equation}
\label{eq:U_rep}
U_{\rm rep} = \sum_{i < j} A^{ij} e^{-B^{ij} r_{ij}}.
\end{equation}
In addition, instead of $U_{\rm disp} + U_{\rm rep}$ in
Eqs.~\eqref{eq:U_disp} and \eqref{eq:U_rep}, the Lennard-Jones potential $U_{\rm LJ}$ 
is used to describe the water-water repulsion and dispersion
in the Dang-Chang water model:
\begin{equation}
\label{eq:U_LJ}
U_{\rm LJ} = \sum_{i < j} 4 \epsilon_{ij} \left[
\left( \frac{\sigma_{ij}}{r_{ij}} \right)^{12} - 
\left( \frac{\sigma_{ij}}{r_{ij}} \right)^6
\right].
\end{equation}
Finally, the polarization energy $U_{\rm pol}$ 
describes electrostatic many-body effects 
that are included through induced dipoles 
${\boldsymbol{\mu}}^i$:
\begin{equation}
\label{eq:U_pol}
U_{\rm pol} = \sum_i \frac{|{\boldsymbol{\mu}}^i|^2}{2 \alpha^i}
+ 
\sum_{i,j} \left[
\left(q^i \mu^j_{\alpha} g^{ij} (r_{ij}) - 
q^j \mu^i_{\alpha} g^{ji} (r_{ij}) \right) T_{ij}^{\alpha} 
- \mu_{\alpha}^i \mu_{\beta}^j T_{ij}^{\alpha\beta}  
\right],
\end{equation}
where Einstein summation is used for Cartesian
components $\alpha$ and $\beta$,
 $\alpha^i$ is the polarizability of the $i$-th atom,
$T_{ij}^\alpha = \partial_{\alpha} (1/r_{ij})$ 
and $T_{ij}^{\alpha\beta} = \partial_{\alpha} \partial_{\beta}
(1/r_{ij})$ are multipole interaction tensors, and 
$g^{ij}(r_{ij}) = 1 - c^{ij} \exp\left({-b^{ij} r_{ij}}\right) 
\sum_{k = 0}^{4} \frac{\left(b^{ij} r_{ij}\right)^k}{k!}$ is a short-range Tang-Toennies correction. The induced dipoles 
are determined at every integration time step through 
minimization of $U_{\rm pol}$ in Eq.~\eqref{eq:U_pol}. In practice,
the induced dipoles for cations in the PIM are neglected due 
to their quite small polarizability\cite{PIM}.
The induced dipole of a Dang-Chang water molecule
is carried by the virtual M-site.
All ion-ion,
ion-water, and water-water interactions parameters in Eqs.~\eqref{eq:U_charge}, \eqref{eq:U_disp}, \eqref{eq:U_rep},
\eqref{eq:U_LJ}, and \eqref{eq:U_pol} were taken from Ref.~\citenum{PIM}. 

The Madrid-2019\cite{Madrid_2019} model for ions in 
aqueous solutions employs only $U_{\rm charge}$ (see Eq.~\eqref{eq:U_charge})
and $U_{\rm LJ}$ (see Eq.~\eqref{eq:U_LJ}) to describe all interactions in
the system:
\begin{equation}
U_{\rm tot}^{\rm Madrid-2019} = U_{\rm charge} + U_{\rm LJ},
\end{equation}
where ions carry scaled charges (+0.85, -0.85, +1.70 for monovalent cations, monovalent anions and divalent cations, respectively) as an effective way to introduce the electronic polarizability of the medium, and sites on water molecules carry partial charges ($q_{\rm H} = +0.5564$, $q_{\rm O} = 0$,  $q_{\rm M} = -2q_{\rm H}$). The corresponding Lennard-Jones
interaction parameters $\epsilon_{ij}$ and $\sigma_{ij}$
can be found in Ref.~\citenum{Madrid_2019}. Finally,
the SPC/E-based Amber14 FF \cite{Joung2008,Joung2009,Li2013,AMBER14} is similarly 
described with
\begin{equation}
U_{\rm tot}^{\rm Amber14} = U_{\rm charge} + U_{\rm LJ}
\end{equation}
with formal ionic charges (+1, -1, +2) and the
partial charges of the SPC/E water molecules ($q_{\rm H} = +0.4238$,
$q_{\rm O} = -2 q_{\rm H}$). The corresponding interaction
parameters were taken from Refs.~\citenum{Joung2008,Joung2009,Li2013}.

To mimic infinite dilution conditions, $N = 256$ water molecules
and 1 ion were simulated in a cubic box at density $\rho=0.997$ g/cm$^3$ and temperature $T=298.15$ K that was maintained 
using the Nosé-Hoover chains thermostat\cite{Nose1984,Hoover1985,NHchains} with a time constant of 1 ps.
Electrostatic interactions were treated using Ewald summation\cite{AguadoMadden, Laino2003}.
The cutoff for short-range interactions was set to
the half of the box side length (close to 1 nm).
Initially, arbitrary-oriented
water molecules and an ion were randomly initialized on a lattice, then annealed at $T = 1000$ K for 150 ps and subsequently equilibrated at $T = 298.15$ K for another 150 ps. 
Such preparation protocol was followed by
5 independent production runs each of length 1 ns in the $NVT$ ensemble  with an integration time step of 1 fs. The water molecules were made effectively rigid by integrating their equations of motion
using the RATTLE\cite{Rattle} algorithm with precision 
$10^{-9}$. All classical MD simulations as well as the EFG
calculations were performed with the MetalWalls package\cite{MW2}.
In the case of PIM simulations, the induced dipoles 
were determined self-consistently using the conjugate gradient 
algorithm with tolerance $10^{-6}$. The computation of the EFG at the ion position 
due to point charges and point dipoles, with the latter contributing
only in the case of PIM simulations, was implemented using Ewald summation\cite{AguadoMadden, Laino2003} 
in the MetalWalls package (release version 21.06\cite{MW_2106}) and 
was performed at every integration time step. 
Finally, it is important to note that in the case of a 
non-electroneutral system, the trace of the classical EFG computed 
using the Ewald summation expressions\cite{AguadoMadden}
features a non-zero trace 
\begin{equation}
\label{eq:efg_trace}
\mathrm{Tr} \, \mathbf{V}_{\rm ext} = - \frac{4 \pi}{V} q_{\rm tot},
\end{equation}
where $q_{\rm tot}$ is the total charge of the simulation box
and $V$ is its volume. To ensure that the final EFG
is traceless, a third of \eqref{eq:efg_trace} was subtracted
from the diagonal EFG components before analysis. In addition, we verified that
the resulting EFG ACFs are almost identical to those in a dilute, explicitly electroneutral
system (see Supporting Fig. S1{\bf a} and S1{\bf b}). Finally, the EFG ACFs obtained from $NVE$ production runs are in excellent agreement with the ones in the $NVT$ ensemble 
(see Supporting Fig. S1{\bf c} and S1{\bf d}).

To determine effective Sternheimer factors, 
for each classical FF we additionally simulated a smaller system
comprised of 64 water molecules and 1 ion using the same system
parameters as discussed before.
The smaller systems were prepared with the same protocol as 
the larger ones. Subsequently, we sampled 1000 configurations every 10 ps during 
a single $NVT$ production run
to ensure a sufficient degree of decorrelation between two 
consecutive snapshots. These 1000 configurations from classical
MD were used as an input for the DFT-based computations of the
EFG in the condensed phase that include the electronic contribution.
The DFT calculations were performed in the Quantum Espresso (QE)
package\cite{Giannozzi_2009}
using the projected augmented wave (PAW) method to allow for an 
all-electron representation of the core region.
\cite{Bloechl1994,Petrilli1998,Charpentier2011}.
The AI EFG at the ion 
position was evaluated with periodic boundary conditions using the QE-GIPAW package\cite{Varini2013}. 
The self-consistent electron densities were obtained using 
the PBE functional\cite{PBE1996} 
and a kinetic energy cutoff of 80 Ry.
In the case of Li$^+$, Na$^+$, K$^+$, Cl$^-$, Mg$^{2+}$,
Ca$^{2+}$ ions we employed the norm-conserving pseudopotentials
included in the GIPAW package\cite{GIPAW_pseudo}, whereas for 
Cs$^+$, Br$^-$, and I$^-$ we used the PAW pseudopotentials
from the pslibrary 1.0.0\cite{pslibrary}.
Due to approximations involved in representing AI EFGs in the present approach,
here we used the PBE functional for all ions, although more advanced functionals can provide a more accurate description of EFGs for larger and more polarizable species \cite{Autschbach2010, Philips2020}. In other words, the errors produced by the 
Sternheimer approximation \eqref{eq:st_approx_eq} might considerably outweigh the ones
associated with the use of a simpler DFT functional.

\begin{figure*}[t!]
\centering
\includegraphics[width=0.99\textwidth]{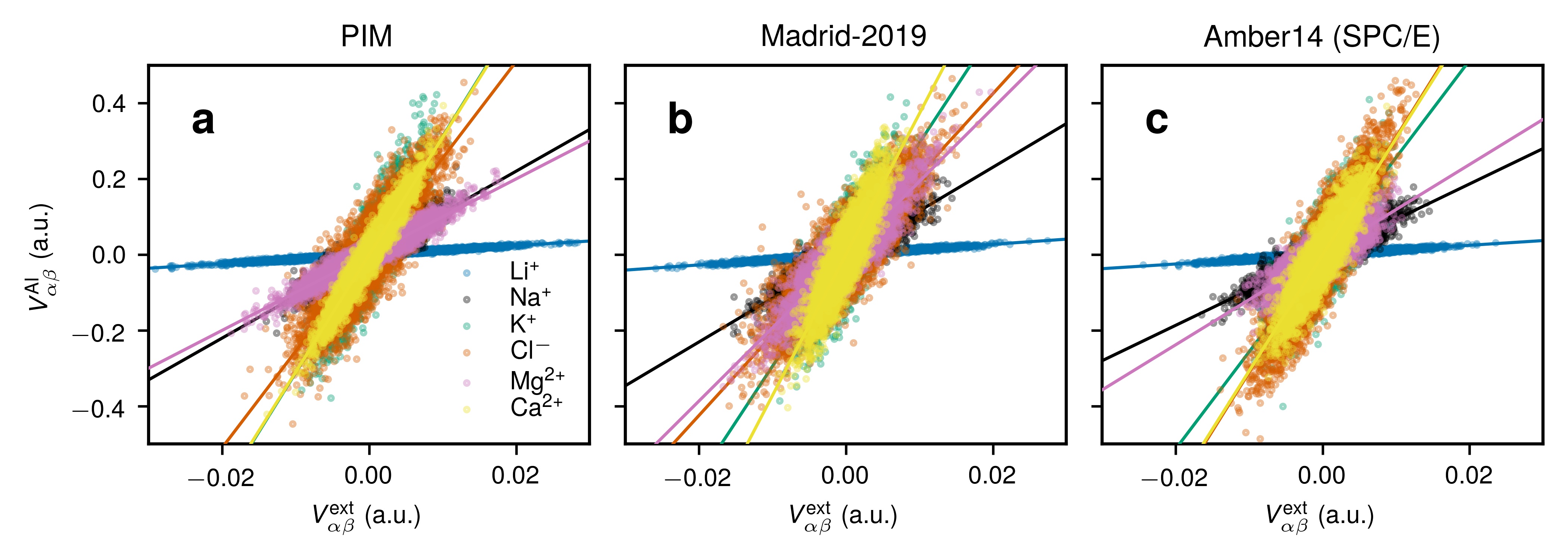}
\caption{\label{fig:sh_approx}
Validation of the Sternheimer approximation
for different classical FFs. 
The EFG components obtained using classical MD, $V_{\alpha\beta}^{\rm ext}$, are plotted versus the ab initio ones,
$V_{\alpha\beta}^{\rm AI}$, following the procedure 
explained in Section~\ref{sec:sim_details}.
A linear relationship
between $V_{\alpha\beta}^{\rm AI}$ and $V_{\alpha\beta}^{\rm ext}$
is evident in all cases considered and allows to define 
an effective Sternheimer factor 
$V_{\alpha\beta}^{\rm AI} = (1 + \gamma_{\rm eff}) 
V_{\alpha\beta}^{\rm ext}$ for every model. The fits for $\gamma_{\rm eff}$ are shown with solid lines, whereas the obtained values 
of $\gamma_{\rm eff}$ are listed in Tab.~\ref{tab:gamma_eff}.
The results for Li$^+$, Na$^+$, K$^+$, Cl$^-$, Mg$^{2+}$, 
and Ca$^{2+}$ ions are indicated with a different color shown in the legend 
of ({\bf a}) (the legend is the same for all plots).
Additional results for Cs$^+$, Br$^-$, and I$^-$ are shown in Supporting Fig.~S2.}
\end{figure*}

%%%%%%%%%%%%%%%%%%%%%%%%%%%%%%%%%%%%%%%%%%%%%%%%%%%%%%%%%%%%%%%%%%%%%%%%%%%%%%%%%%
\section{Sternheimer approximation}
\label{sec:stern_approx}

%%%%%%%%%%%
\subsection{Effective Sternheimer factor}

\begin{figure}[t!]
\centering
\includegraphics[width=0.6\linewidth]{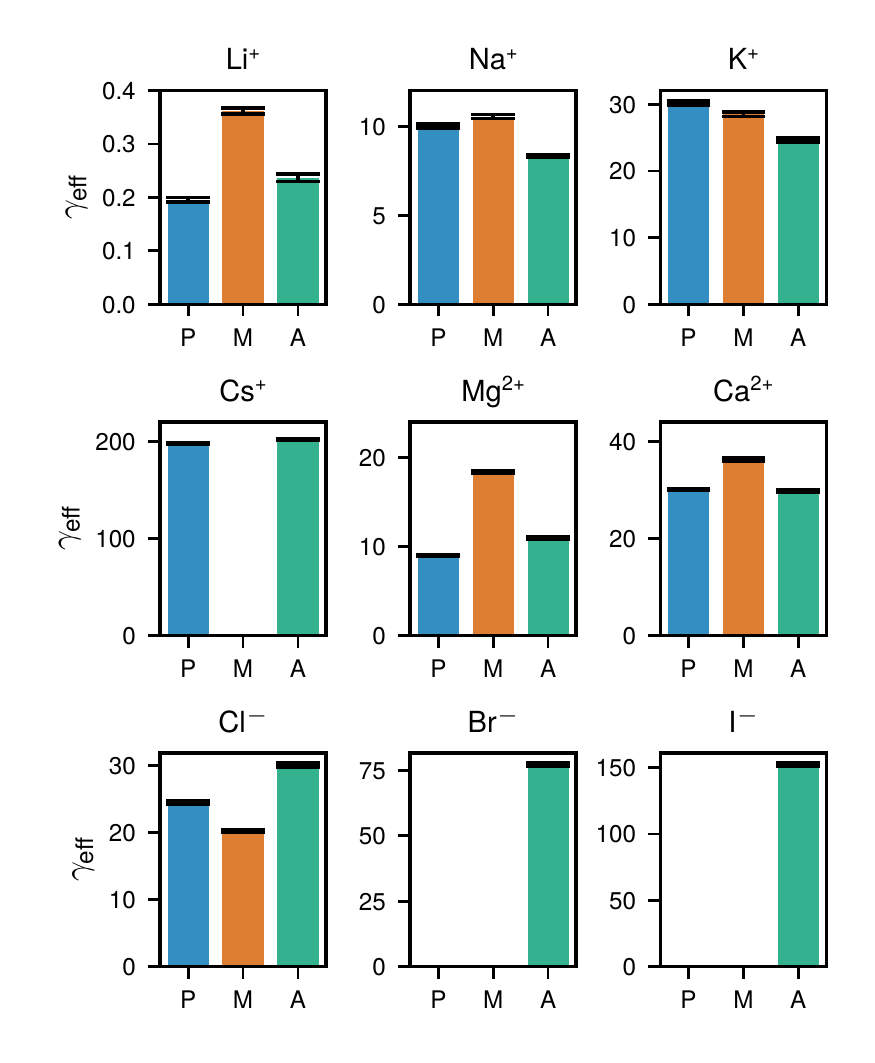}
\caption{\label{fig:gamma_eff}
The effective Sternheimer factors for 
different quadrupolar nuclei at infinite dilution. For every ion,
we indicate $\gamma_{\rm eff}$ for the PIM (P),
the Madrid-2019 model (M),
and Amber14 FF parameters for the SPC/E water and ions (A).}
\end{figure}

We first assess the validity of the Sternheimer approximation
by comparing the EFGs obtained with classical MD simulations,
$V_{\alpha\beta}^{\rm ext}$, to that from DFT GIPAW calculations,
$V_{\alpha\beta}^{\rm AI}$, on a set of 1000 configurations.
The results for Li$^+$, Na$^+$, K$^+$, Cl$^-$, Mg$^{2+}$, 
and Ca$^{2+}$ ions, the parameters for which are available
in all three FFs considered in this work, are shown in Fig.~\ref{fig:sh_approx}. Additional results for Cs$^+$, 
Br$^-$, I$^-$, and Ca$^{2+}$ ions can be found in Supporting Figs.~S2 and S3.
In agreement with earlier work\cite{Carof2014},
we find that the linear relation $V_{\alpha\beta}^{\rm AI} 
\simeq (1 + \gamma_{\rm eff}) V_{\alpha\beta}^{\rm ext}$ 
holds well in all of the cases considered (that is, for all
ions and FFs) and allows to determine effective model-dependent
Sternheimer factors $\gamma_{\rm eff}$ using a linear fit
\eqref{eq:st_approx_eq}. 
Note that here and in what follows we will distinguish between
the effective Sternheimer factors $\gamma_{\rm eff}$ derived for
a specific atomic environment and $\gamma_{\rm \infty}$ that
are calculated using perturbation methods for model
condensed phase environments or free ions in response to a distant 
charge distribution\cite{Sternheimer1966, Schmidt1980}.
From the slope of the scatter plots in Fig.~\ref{fig:sh_approx} and Supporting Fig.~S2, we consistently find that $V_{\alpha\beta}^{\rm AI}$ responds positively to the EFG of the external (ionic) charge distribution, $V_{\alpha\beta}^{\rm ext}$, in the present approach,
leading to the anti-shielding effect and strictly positive values of the obtained Sternheimer factors. This differs from the case of $\gamma_{\infty}$ for Li$^+$ obtained \emph{via} the Watson sphere model, which aimed at modeling a solid-like environment\cite{Schmidt1980}, that results 
in the shielding effect for the cation and a negative value of $\gamma_{\infty}$.
Further below we will compare in more detail the values of $\gamma_{\rm eff}$ devised in this work
and that of $\gamma_{\infty}$ from the Watson sphere model that were used in the early systematic 
study on the NMR relaxation rates of quadrupolar nuclei by Roberts and Schnitker\cite{RS1993}.

\begin{table*}[t!]
\caption{\label{tab:gamma_eff}
Effective Sternheimer factors $\gamma_{\rm eff}$
and corresponding prediction errors $\sigma(\mathbf{V}^2)$ on the EFG variance (see Eq.~\eqref{eq:error}) obtained for different classical FFs. 
The values in parentheses indicate standard errors obtained using bootstrapping. 
The reference values of $\gamma_{\infty}$ correspond 
to the Watson sphere model on an ionic solid\cite{Schmidt1980}. Note that 
here the sign of $\gamma_{\infty}$ is opposite to the one in Ref.~\citenum{Schmidt1980}
to comply with the convention in Eq.~\eqref{eq:st_approx_eq}.}
\vspace*{2mm}
\small
\begin{tabular}{cccccccc}
\toprule
\multirow{2}{*}{Ion} & \multirow{2}{*}{$\gamma_{\infty}$} 
& \multicolumn{2}{c}{PIM} & \multicolumn{2}{c}{Madrid-2019} & \multicolumn{2}{c}{Amber14} \\
\cmidrule(lr){3-4}\cmidrule(lr){5-6}\cmidrule(lr){7-8}
 {} & {}  
 & $\gamma_{\rm eff}$ & $\sigma(\mathbf{V}^2)$ 
 & $\gamma_{\rm eff}$ & $\sigma(\mathbf{V}^2)$ 
 & $\gamma_{\rm eff}$ & $\sigma(\mathbf{V}^2)$ \\ \midrule
Li$^+$ & -0.255 & 0.196(0.004) & 0.20 & 0.362(0.006)   & 0.26 
& 0.237(0.007) & 0.29 \\ \midrule
Na$^+$ & 5.452 & 10.02(0.11) & 0.46 & 10.54(0.11)    & 0.46 & 8.34(0.09) & 0.43 \\ \midrule
K$^+$  & 21.782 & 30.21(0.32) & 0.47 & 28.51(0.33)     & 0.47 & 24.67(0.29) & 0.48 \\ \midrule
Cs$^+$  & 110.81 & 196.7(1.0) & 0.25 & ---     & --- & 201.8(1.0) & 0.26 \\ \midrule
Cl$^-$ & 41.999 & 24.50(0.24) & 0.42 & 20.25(0.22) & 0.42 & 30.06(0.27) & 0.39 \\ \midrule
Br$^-$ & 85.517 & --- & --- & --- & --- & 77.16(0.65) & 0.39 \\ \midrule
I$^-$ & 162.42 & --- & --- & --- & --- & 152.1(1.3) & 0.40 \\ \midrule
Mg$^{2+}$ & 4.118 & 9.00(0.06) & 0.36 & 18.34(0.17) & 0.55 & 10.91(0.15) & 0.56 \\ \midrule
Ca$^{2+}$ & 18.791 & 30.02(0.13) & 0.23 & 36.24(0.30) & 0.36 & 29.73(0.23) & 0.36 \\ \bottomrule
\end{tabular}
\end{table*}

The comparison between $\gamma_{\rm eff}$ for different classical
FFs are shown in Fig.~\ref{fig:gamma_eff}, whereas explicit
values are listed in Tab.~\ref{tab:gamma_eff} (the corresponding
standard errors were estimated using bootstrapping).
As expected, for the considered atomic species the effective Sternheimer factor 
increases with the number of electrons, indicating its dominant 
contribution to the total EFG at the nucleus position.
In addition, certain quantitative differences are found when comparing $\gamma_{\rm eff}$ across the three classical FFs considered,
highlighting the sensitivity of local polarization effects
to the charge distribution around an ion and solute-solvent
interactions. For the two smallest 
cations, Li$^+$ and Mg$^{2+}$, $\gamma_{\rm eff}$ in the Madrid-2019
model are larger by a factor $\approx$ 1.5--2 than those in the PIM and Amber14 FFs.
A similar trend is found for Ca$^{2+}$, however the magnitude of 
$\gamma_{\rm eff}^{\rm Madrid-2019}$ is bigger by a factor $\approx$ 1.2 
when compared to the two other FFs. Smaller differences in
$\gamma_{\rm eff}$ across models are seen for larger
Na$^+$, K$^{+}$, and Cs$^+$ cations. 
For the Cl$^-$ anion,
$\gamma_{\rm eff}$ for the Amber14 FF is larger by a factor  $\approx$ 1.25--1.5 than
those for the PIM and Madrid-2019 model.

When the comparison is made across different FFs, 
two effects may play an important role: $(i)$ the representation of the charge density around an ion using different (partial) point charges and point dipoles (in polarizable models) on water molecules; $(ii)$ changes in the structure and dynamics of 
the solvation shell around an ion. Both effects can potentially influence $V_{\alpha\beta}^{\rm AI}$ and $V_{\alpha\beta}^{\rm ext}$, 
and therefore the computed value of $\gamma_{\rm eff}$.
To assess the effect of charge density representation via distinct water 
point charges and dipoles, we have adopted the following strategy. First, we took
1000 system configurations that were used in the AI calculations of the 
Sternheimer factor for the PIM and recomputed the external EFGs using
the point charges from the Madrid-2019 FF (this is possible thanks to the same TIP4P/2005 geometry of water molecules in these two FFs). The latter allowed us to calculate 
Sternheimer factors for configurations that were generated with the PIM dynamics 
but with point charges of the Madrid-2019 FF (Supporting Fig.~S4). Interestingly,
we find that such operation results in quite larger values of $\gamma_{\rm eff}$ 
for Li$^+$, Mg$^{2+}$, Ca$^{2+}$ that become closer to the ones obtained consistently 
for the Madrid-2019 FF (Tab.~\ref{tab:gamma_eff}). Yet, the values of $\gamma_{\rm eff}$ 
obtained after such a numerical experiment are not in quantitative agreement with 
those for the Madrid-2019 FF (Tab.~\ref{tab:gamma_eff}), suggesting additional 
differences that stem from variations in the solvation shell structure in the 
two models at hand, which are highlighted in the difference in the ion-oxygen radial distribution functions and in the number of water molecules in the shell (Supporting Figs. S6 and S7, respectively).
Furthermore, much smaller variations in $\gamma_{\rm eff}$ are found for Na$^+$ and K$^{+}$.
In addition, $\gamma_{\rm eff}$ for Cl$^{-}$ increases contrary to the corresponding value in the Madrid-2019 FF (Tab.~\ref{tab:gamma_eff}). Consistently, opposite trends for 
$\gamma_{\rm eff}$ of Li$^+$, Mg$^{2+}$, Ca$^{2+}$, and Cl$^{-}$ are observed when 
the Sternheimer factors are calculated using configurations that were generated with 
the Madrid-2019 dynamics but with point charges and dipoles of the PIM (Supporting Fig.~S5). In summary, the latter observations highlight the 
sensitivity of the obtained $\gamma_{\rm eff}$ to the charge density representation of the chosen water model. Further differences in $\gamma_{\rm eff}$ are likely caused by subtle variations 
in the solvation shell structure across the FFs (Supporting Figs.~S6 and S7).
For instance, the Amber14 FF features approximately one more molecule in the first solvation shell of the Cl$^-$ anion on average as compared to 
the two other models (6.9 vs. 6.2 and 5.9, respectively, see Supporting Fig.~S7 and Tab.~S1), which can result in an enhanced polarization of the electronic cloud by hydrogen bonds donated by water for such an environment. 
Finally, the observed deviations in computed $\gamma_{\rm eff}$
across different classical FFs again emphasize the sensitivity
of the EFGs to the local environment around an ion. 

The liquid-state environment around a solute impacts the 
obtained Sternheimer factors $\gamma_{\rm eff}$, as compared to those from the Watson sphere model $\gamma_{\rm \infty}$, 
in which the ion's electron
cloud responds to the surrounding, oppositely charged hollow sphere 
that models the environment of an ionic solid \cite{Schmidt1980}. 
Here we briefly discuss 
the differences between $\gamma_{\rm eff}$ and 
$\gamma_{\rm \infty}$ (see Fig.~\ref{fig:sh_approx} and Tab.~\ref{tab:gamma_eff}), 
as the latter were used in the early
systematic study of quadrupolar relaxation rates\cite{RS1993}.
For the least polarizable cation, Li$^+$, the effective Sternheimer 
factors are quite small ($\gamma_{\rm eff} \approx$ 0.2--0.36),
yet of different sign when compared to $\gamma_{\infty} = -0.255$.
Such discrepancy might be due 
to differences in the scaled-charge model and 
ab initio solvent charge distributions around an ion, 
as highlighted in Ref.~\citenum{Kathmann2011}.
For alkali metal cations, we find that 
$\gamma_{\rm eff}$ is generally larger than 
$\gamma_{\rm \infty}$ (approximately by a factor of 2 for  
Na$^+$ and Cs$^+$, and by a factor of 1.1--1.4 for K$^+$).
For halide ions, $\gamma_{\rm eff}$ is 
somewhat smaller than $\gamma_{\rm \infty}$
(smaller by a factor of $\approx$ 1.1 for Br$^-$ and I$^-$, and by a factor of 1.4--2
for Cl$^-$ depending on the classical
FF considered). For Mg$^{2+}$ and Ca$^{2+}$, 
$\gamma_{\rm eff}$ is 
around 2--4.5 and 1.6--2 times larger than $\gamma_{\rm \infty}$, 
respectively.
In addition, the Sternheimer-like polarization factors have recently
been obtained from analyzing contributions of localized orbitals
to the EFG for large and highly polarizable ions in AIMD\cite{Philips2020}. The ratio between the total and
external EFG in Ref.~\citenum{Philips2020} was found to be
+3 for Cs$^+$ and -65 for I$^-$, differing considerably 
both from $\gamma_{\rm \infty}$ and also $\gamma_{\rm eff}$ obtained in this work. Yet, as highlighted in Ref.~\citenum{Philips2020}, 
a clean separation between external and internal EFGs is not 
always meaningful in systems with a degree of donation or hydrogen bonding.

\begin{figure}[t!]
\centering
\includegraphics[width=0.75\linewidth]{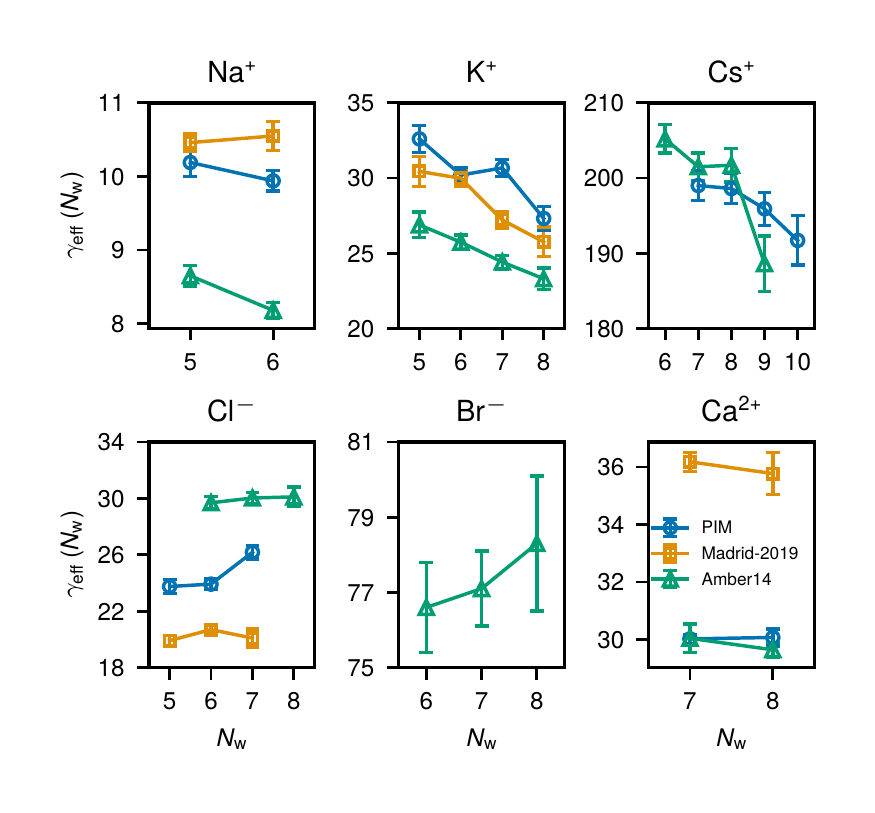}
\caption{\label{fig:gamma_eff_nw}
Effective Sternheimer factors for 
different quadrupolar ions at infinite dilution as 
a function of the number of water molecules $N_{\rm w}$ 
in the first hydration shell. We compare $\gamma_{\rm eff}$ obtained with the charge distribution in the PIM (blue circles),
the Madrid-2019 model (yellow squares),
and Amber14 FF parameters for the SPC/E water and ions (green triangles).}
\end{figure}

We now assess the quality of the Sternheimer approximation
\begin{equation}
\label{eq:SA}
V_{\alpha\beta}^{\rm SA} = (1 + \gamma_{\rm eff}) V_{\alpha\beta}^{\rm ext}
\end{equation}
by considering the average prediction error for the squared 
EFG:
\begin{equation}
\label{eq:error}
\sigma(\mathbf{V}^2) = \frac{1}{M} \sum_{i = 1}^{M}
\left| 
\frac{\left(\mathbf{V}^{\rm SA}_i \right)^2 - 
\left(\mathbf{V}^{\rm AI}_i \right)^2}
{\left(\mathbf{V}^{\rm AI}_i \right)^2}
\right|,
\end{equation}
where $M = 1000$ is the total number of configurations
used for analysis, and $\mathbf{V}_i^2$ above stands for 
$\mathbf{V}_i:\mathbf{V}_i$ as computed from the Sternheimer 
approximation \eqref{eq:SA}, $\left(\mathbf{V}^{\rm SA}_i \right)^2$, or from DFT GIPAW calculations, $\left(\mathbf{V}^{\rm AI}_i \right)^2$, for the same $i$-th configuration. The 
values of $\sigma(\mathbf{V}^2)$ are listed in Tab.~\ref{tab:gamma_eff}.
Generally, for every ion and classical FF, we find $\sigma(\mathbf{V}^2)$
to be quite large, indicating a 20-50\% percent difference 
between   $\left(\mathbf{V}^{\rm SA} \right)^2$ and
$\left(\mathbf{V}^{\rm AI} \right)^2$ on average. Since such 
relatively large prediction error might considerably impact
the resulting NMR relaxation rates that are directly proportional
to the EFG variance $\left\langle \mathbf{V}^2 \right\rangle$, 
in what follows we attempt to construct a better approximation
for the EFG beyond the linear regime \eqref{eq:SA}.

\subsection{Effect of the first solvation shell}

Since the main contribution to the EFG at the ion position arises from the closest solvent molecules\cite{Carof2015},
we start with investigating the potential impact of the hydration shell structure on the obtained effective Sternheimer factors.
Fig.~\ref{fig:gamma_eff_nw} shows $\gamma_{\rm eff}$ in the three
considered classical FFs as a function of the number of water
molecules $N_{\rm w}$ in the first hydration shell of an ion for 
Na$^+$, K$^+$, Cl$^-$, Br$^-$, and Ca$^{2+}$ (see also Supporting
Tab.~S2). The first minimum of the 
ion-oxygen radial distribution function (RDF) was used as a boundary of the first solvation shell (see Supporting~Fig.~S6
and Supporting~Tab.~S1). As Li$^+$ and Mg$^{2+}$ feature a quite
stable hydration shell structure (Supporting~Fig.~S7) with
$N_{\rm w} = 4$ and 6, respectively, they were
excluded from the present analysis. Among the 1000
configurations used for the AI EFG parametrization, for a given
ion and FF we selected only those that had at least 50 occurrences
for a specific value of $N_{\rm w}$. The corresponding standard
errors in Fig.~\ref{fig:gamma_eff_nw} and Supporting Tab.~S2 were also estimated using bootstrapping. 

For most of the cations, $\gamma_{\rm eff}$ decreases with $N_{\rm w}$, yet enhanced statistics is necessary to decisively confirm this trend. Relatively small differences in $\gamma_{\rm eff}(N_{\rm w})$ (up to around 5\%) are found for the Na$^+$ and Ca$^{2+}$ cations 
when comparing their two most probable coordination numbers 
(5 and 6 for Na$^+$; 7 and 8 for Ca$^{2+}$). The K$^+$ cation 
has a quite broad distribution of coordination numbers (Supporting~Fig.~S7) and its $\gamma_{\rm eff}(N_{\rm w})$  features the most pronounced dependence on $N_{\rm w}$ among
the ions considered. In particular, consistently across the 
three FFs, $\gamma_{\rm eff}$ for K$^+$ 
decreases by up to $\approx15$\%
when going from 5 to 8 water molecules in the first solvation
shell. An up to 10\% reduction in $\gamma_{\rm eff}$ is found 
for Cs$^+$ for increasing $N_{\rm w}$ in the PIM and the Amber14 FF.
For Cl$^-$, $\gamma_{\rm eff}$ increases by around 10\% for 
increasing $N_{\rm w}$ from 5 to 7 in the PIM. For anions 
in the Amber14 FF, relatively small changes in the Sternheimer 
factor
are seen with increasing $N_{\rm w}$, yet $\gamma_{\rm eff}$
tends to grow for Br$^-$ and Cl$^-$. In summary, taking into
account the hydration shell structure around an ion might be 
necessary to improve the prediction for the final relaxation 
rate (especially, in such cases as K$^+$ and Cs$^+$ for which
$\gamma_{\rm eff}$ features a relatively pronounced 
dependence on $N_{\rm w}$). 

%%%%%%%%%%%
\subsection{Beyond the standard Sternheimer approximation}

\begin{figure}[t!]
\centering
\includegraphics[width=0.8\linewidth]{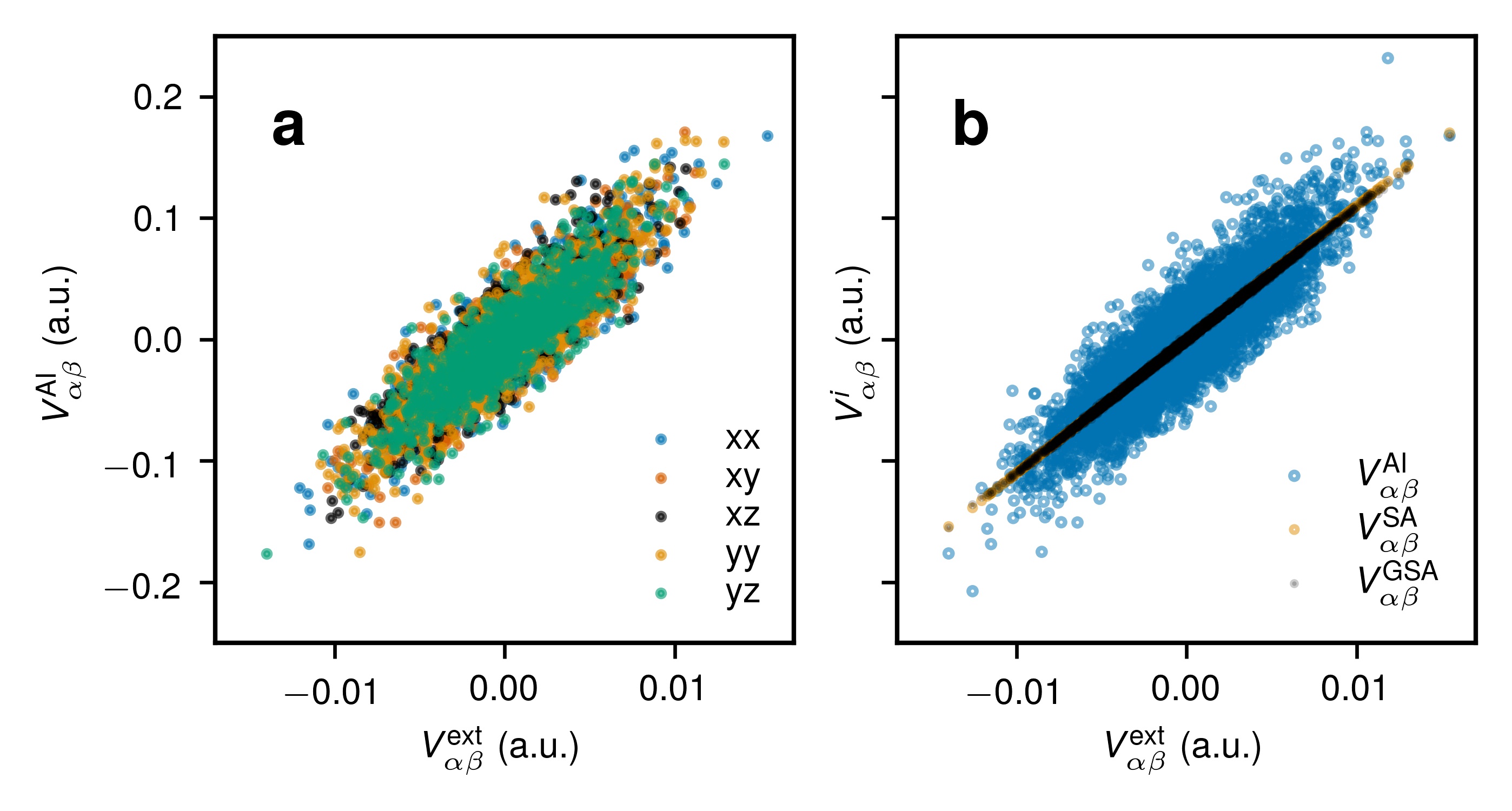}
\caption{\label{fig:gsa}
({\bf a}) Distribution of individual components of 
$V_{\alpha\beta}^{\rm AI}$ plotted versus
$V_{\alpha\beta}^{\rm ext}$.
({\bf b}), The original distribution of $V_{\alpha\beta}^{\rm AI}$ 
as a function of $V_{\alpha\beta}^{\rm ext}$ is compared to 
$V_{\alpha\beta}^{\rm SA}$ obtained using the linear Sternheimer approximation \eqref{eq:SA} and its generalized form \eqref{eq:GSA}. The EFGs shown here correspond to the Na$^+$ ion
in the PIM. Qualitatively similar dependencies are found 
for other ions and FFs.} 
\end{figure}

To improve the quality of the original Sternheimer approximation 
\eqref{eq:SA} and to capture the differences in $\gamma_{\rm eff}$ generated by 
a distinct number of water molecules in the first solvation
shell, we have attempted to construct a generalized linear
Sternheimer approximation (GSA) that introduces 
couplings between diagonal and off-diagonal components of 
$V_{\alpha\beta}^{\rm ext}$. Although the scatter plots
of $V^{\rm AI}_{\alpha\beta}$ plotted versus $V^{\rm ext}_{\alpha\beta}$ are practically identical for both diagonal and off-diagonal components (see Fig.~\ref{fig:gsa}{\bf a}), 
yielding very similar component-wise Sternheimer factors, such GSA might be useful
for capturing the differences in the electronic environments 
of an ion for each specific $N_{\rm w}$. Specifically, to parametrize
the GSA we fit $V^{\rm AI}_{\alpha\beta}$ against $V_{\alpha\beta}^{\rm ext}$ employing the following expression:
\begin{equation}
\label{eq:GSA}
V^{\rm GSA}_{\alpha\beta} = (1 + \gamma_1)
V_{\alpha\beta}^{\rm ext} + 
\gamma_2 \Gamma_{\alpha \gamma} V_{\gamma\delta}^{\rm ext} \Gamma^T_{\delta\beta},
\end{equation}
where $\gamma_1$ and $\gamma_2$ are two Sternheimer-like factors,
and $\Gamma_{\rm \alpha\beta}$ is an orthonormal matrix 
that generates an additional coupling between the components
of $V_{\alpha\beta}^{\rm ext}$
and satisfies $\Gamma_{\alpha \gamma} \Gamma^T_{\gamma\beta} = \delta_{\alpha\beta}$. Note that by construction the term 
$\Gamma_{\alpha \gamma} V_{\gamma\delta}^{\rm ext} \Gamma^T_{\delta\beta}$ is traceless and symmetric. To perform
the fit \eqref{eq:GSA}, we expressed $\Gamma_{\alpha\beta}$ 
using the Rodriguez formula:
\begin{equation}
\Gamma_{\alpha\beta} (\mathbf{\hat{e}}, \psi) = 
(\cos\psi) \, \delta_{\alpha \beta} + (1 - \cos\psi) \, \hat{e}_\alpha \hat{e}_\beta - (\sin\psi) \, \epsilon_{\alpha\beta\gamma} \hat{e}_{\gamma},
\end{equation}
where $\epsilon_{\alpha\beta\gamma}$ is the Levi-Civita symbol and 
$\mathbf{\hat{e}} = (\sin\theta \cos\phi, \sin\theta \sin\phi, \cos\theta)^T$ is a unit vector. Thus, the angles 
$\psi, \theta, \phi$ and the two constants $\gamma_1, \gamma_2$ 
are the fit parameters for the model \eqref{eq:GSA}. To assess the 
fit quality, in Fig.~\ref{fig:gsa}{\bf{b}} we show $V_{\alpha\beta}^{\rm AI}$, $V_{\alpha\beta}^{\rm SA}$, and $V_{\alpha\beta}^{\rm GSA}$ plotted  versus $V_{\alpha\beta}^{\rm ext}$  
for the Na$^+$ ion in the PIM (qualitatively similar results are obtained for other ions in all FFs considered). It is evident 
that the main error in $V_{\alpha\beta}^{\rm SA}$ comes from 
the failure of the simple linear approximation \eqref{eq:SA} to
capture large off-diagonal deviations of $V^{\rm AI}_{\alpha\beta}$ 
when compared to $V^{\rm ext}_{\alpha\beta}$. Despite a more 
advanced structure of the GSA \eqref{eq:GSA} with 4 additional
fit parameters, $V_{\alpha\beta}^{\rm GSA}$ does not bring a considerable improvement
to the resulting EFG variance (generally, a few percent reduction 
in $\sigma(\mathbf{V}^2)$) and, similarly to $V_{\alpha\beta}^{\rm SA}$,
it fails to capture significant off-diagonal deviations. 
In addition, we find that $\gamma_1$ in the fit \eqref{eq:GSA}
is quite close to $\gamma_{\rm eff}$ in \eqref{eq:SA},
thus making the contribution of $\gamma_2 \Gamma_{\alpha \gamma} V_{\gamma\delta}^{\rm ext} \Gamma^T_{\delta\beta}$ generally small.

\begin{figure}[t!]
\centering
\includegraphics[width=0.75\linewidth]{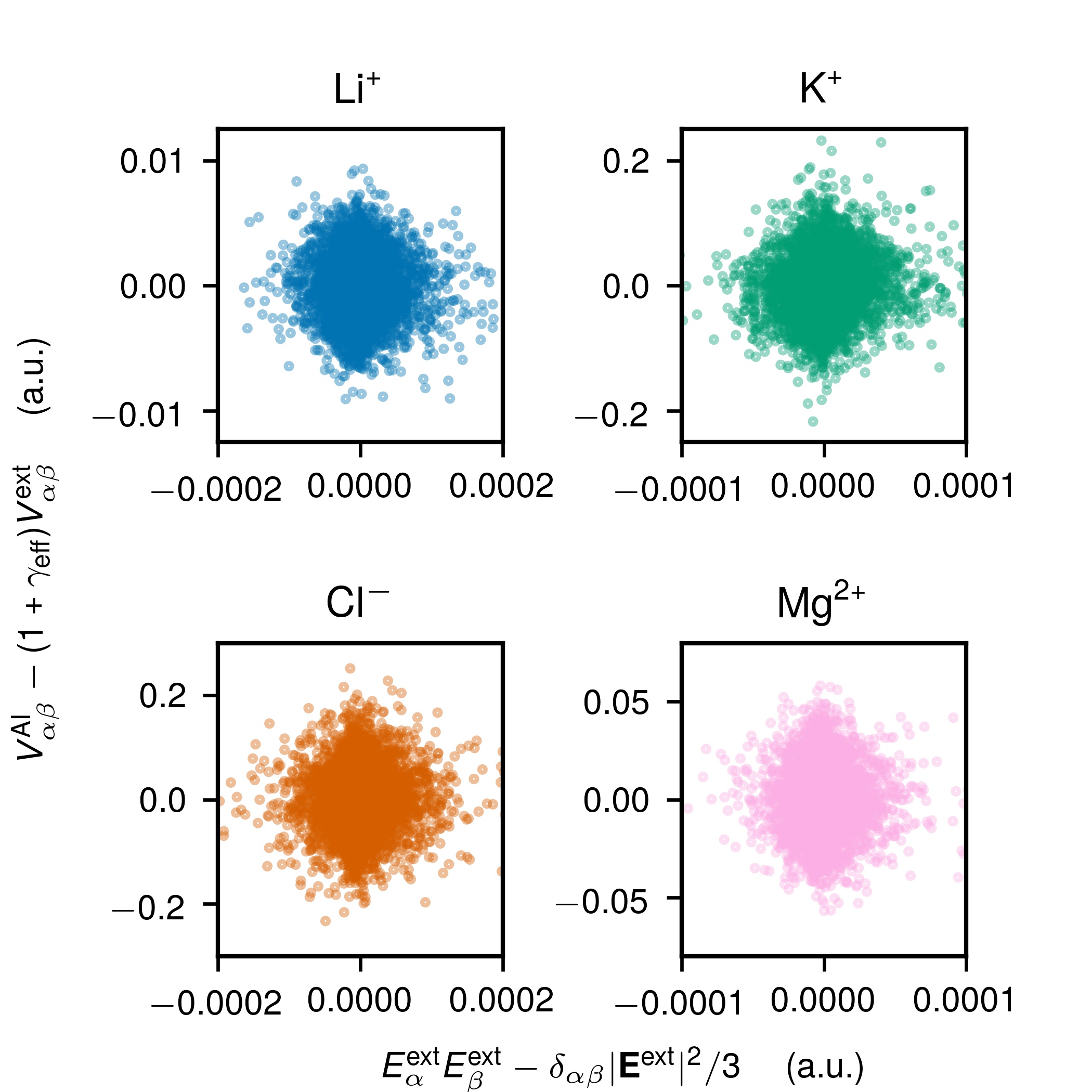}
\caption{\label{fig:nonlin_stern}
The difference between $V_{\alpha\beta}^{\rm AI}$
and $(1 + \gamma_{\rm eff}) V_{\alpha\beta}^{\rm ext}$ is 
plotted versus the traceless version of the
tensor $E_{\alpha}^{\rm ext} E_{\beta}^{\rm ext}$ to account for non-linear effects \eqref{eq:efg_general2} for different 
ions in the PIM. The absence of linear correlation is evident in all cases considered (also for other ions and FFs).} 
\end{figure}

As the GSA \eqref{eq:GSA} had not brought a considerable improvement in predicting the EFG variance compared to the 
AI data, we attempted 
to include the first-order non-linear correction to the 
EFG at the nucleus.
In general, the EFG at the nucleus of an ion $V_{\alpha\beta}$ subject to 
external electric fields can be expressed as a perturbation 
series in the field and its gradients\cite{Fowler1989,Calandra2002}:
\begin{equation}
\label{eq:efg_general}
V_{\alpha\beta} = V_{\alpha\beta}^{\rm ext} + 
g_{\alpha\beta,\gamma} E^{\rm ext}_{\gamma}
+ g_{\alpha\beta,\gamma\delta} V^{\rm ext}_{\gamma\delta}
+ \tfrac{1}{2} \epsilon_{\alpha\beta,\gamma\delta} 
E^{\rm ext}_{\gamma} E^{\rm ext}_{\delta} + \dots,
\end{equation}
where the next order terms above correspond both to higher derivatives of the external field $E_{\alpha}^{\rm ext}$ and higher order non-linear terms.
$g_{\alpha\beta,\gamma}$,  $g_{\alpha\beta,\gamma\delta}$,
and $\epsilon_{\alpha\beta,\gamma\delta}$ are tensorial
susceptibilities that describe the response of the electronic 
cloud in different environments. In the case of an initially spherical electronic cloud of an ion, $g_{\alpha\beta,\gamma} = 0$,
whereas $g_{\alpha\beta,\gamma\delta}$ and 
$\epsilon_{\alpha\beta,\gamma\delta}$ both feature only one 
independent component, $\gamma$ and $\epsilon$, respectively\cite{Calandra2002,Buckingham1967}:
\begin{equation}
\label{eq:gabcd}
g_{\alpha\beta,\gamma\delta} = \gamma \left\lbrace
\frac{1}{2} \left(\delta_{\alpha\gamma} \delta_{\beta\delta}
+ \delta_{\alpha\delta} \delta_{\beta\gamma} \right) 
- \frac{1}{3} \delta_{\alpha\beta} \delta_{\gamma\delta} 
\right \rbrace
\end{equation}
\begin{equation}
\label{eq:eabcd}
\epsilon_{\alpha\beta,\gamma\delta} = \epsilon \left\lbrace
\frac{3}{4} \left(\delta_{\alpha\gamma} \delta_{\beta\delta}
+ \delta_{\alpha\delta} \delta_{\beta\gamma} \right) 
- \frac{1}{2} \delta_{\alpha\beta} \delta_{\gamma\delta} 
\right \rbrace
\end{equation}
Given the relations in Eqs.~\eqref{eq:gabcd}, \eqref{eq:eabcd}
and assuming that the trace of $V_{\alpha\beta}^{\rm ext}$
is identically zero,
the EFG in Eq.~\eqref{eq:efg_general} simplifies to
\begin{equation}
\label{eq:efg_general2}
V_{\alpha\beta} = (1 + \gamma) V_{\alpha\beta}^{\rm ext} + 
\frac{3\epsilon}{4} \left(
E_{\alpha}^{\rm ext} E_{\beta}^{\rm ext}
- \frac{1}{3} \delta_{\alpha\beta} |\mathbf{E}^{\rm ext}|^2  
\right) + \dots,
\end{equation}
where the first term in the equation above corresponds to the usual
Sternheimer approximation and the second one conveys the 
first order non-linear correction with the hyperpolarizability $\epsilon$. In Fig.~\ref{fig:nonlin_stern},
we plot the difference between $V^{\rm AI}_{\alpha\beta}$
and $(1 + \gamma_{\rm eff}) V^{\rm ext}_{\alpha\beta}$ 
versus the traceless tensor ${\widetilde{E}}_{\alpha\beta}^{\rm ext} = E_{\alpha}^{\rm ext} E_{\beta}^{\rm ext}
- \delta_{\alpha\beta} |\mathbf{E}^{\rm ext}|^2 /3$
for Li$^+$, K$^+$, Cl$^-$, and Mg$^{2+}$ in the PIM 
(qualitatively similar results are found for other ions in 
all FFs considered). In agreement with earlier 
results on a smaller number of configurations\cite{CarofPhd}
and in contrast to crystalline atomic environments\cite{Calandra2002}, 
the overall symmetric scatter plots 
centered at the origin indicate the absence of significant linear correlation
between $V^{\rm AI}_{\alpha\beta} - (1 + \gamma_{\rm eff}) V^{\rm ext}_{\alpha\beta}$ and ${\widetilde{E}}_{\alpha\beta}^{\rm ext}$. The fact that the latter two quantities
are uncorrelated in the present liquid state system might be
related to different symmetry properties of the $V_{\alpha\beta}$ and ${\widetilde{E}}_{\alpha\beta}$ tensors. For instance, for certain configurations that feature a symmetry axis the electric field can be vanishing at the origin corresponding to the nucleus
position, whereas the EFG can still remain non-zero. In addition,
we attempted to correlate $V^{\rm AI}_{\alpha\beta} - (1 + \gamma_{\rm eff}) V^{\rm ext}_{\alpha\beta}$ with other higher 
order non-linear terms like $(\mathbf{E}^{\rm ext})^2 V_{\alpha\beta}$, $(\mathbf{V}^{\rm ext})^2 {\widetilde{E}}_{\alpha\beta}^{\rm ext}$, ${\widetilde{E}}_{\alpha\gamma}^{\rm ext} V_{\gamma\beta}$, however, similarly to Fig.~\ref{fig:nonlin_stern},
in all cases no pronounced correlation has been identified (not shown).
Therefore, in the following we use the initial Sternheimer approximation for the prediction of NMR relaxation rates.

%%%%%%%%%%%%%%%%%%%%%%%%%%%%%%%%%%%%%%%%%%%%%%%%%%%%%%%%%%%%%%%%%%%%%%%%%%%%%
\section{NMR relaxation rates}
\label{sec:nmr_rates}

\begin{figure*}[t!]
\centering
\includegraphics[width=0.99\textwidth]{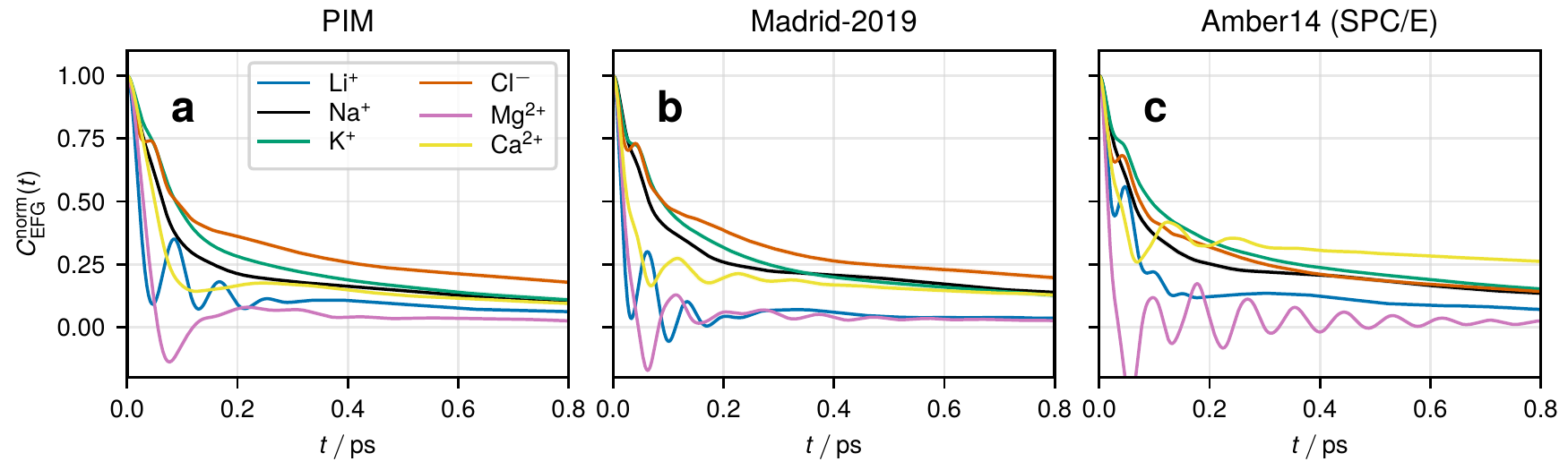}
\vspace*{3mm}
\includegraphics[width=0.98\textwidth]{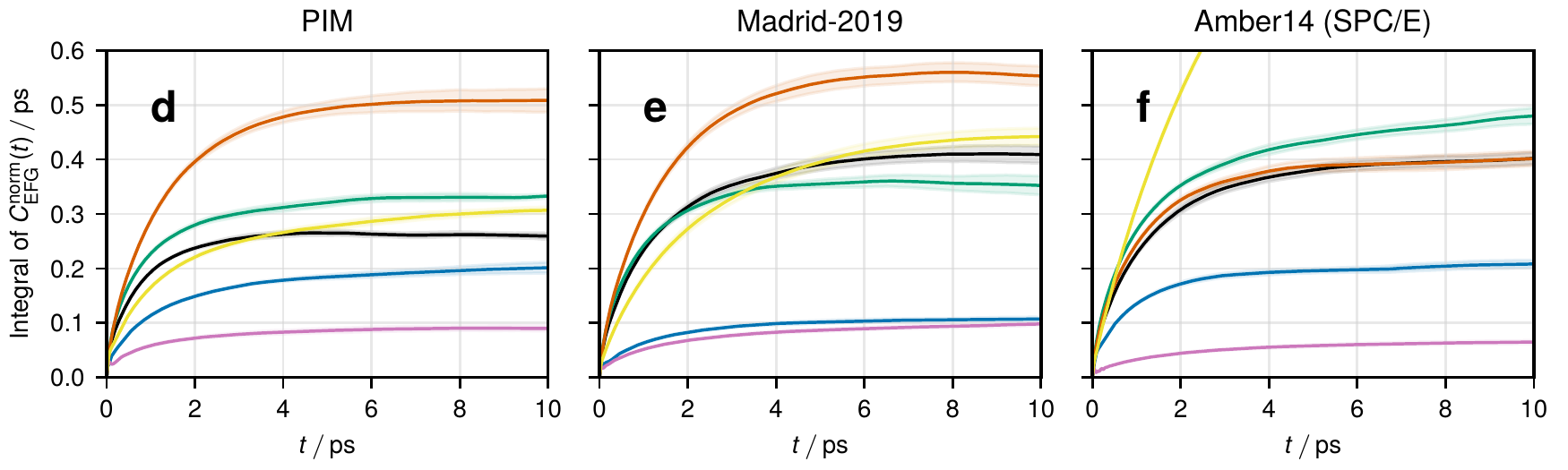}
\caption{\label{fig:efg_acfs} 
Autocorrelation function of the the electric field gradient at the ion position
in different classical FFs: normalized EFG ACF and its running integral in the PIM (({\bf a}) and ({\bf d})), the Madrid-2019 model (({\bf b}) and ({\bf e})),
and Amber14 FF parameters
for the SPC/E water and ions (({\bf c}) and ({\bf f}))
for K$^+$, Na$^+$, Li$^+$, Cl$^-$, and Mg$^{2+}$.
The legend shown in ({\bf a}) also applies to all panels;
panels ({\bf a}), ({\bf b}) and ({\bf c}) share the same range on both axes,
and so do panels ({\bf d}), ({\bf e}) and ({\bf f}).
Shaded regions around a curve correspond to the standard error
obtained from independent simulation runs.
}
\end{figure*}

We finally consider the EFG ACFs from
different classical FFs and the resulting NMR relaxation rates.
Figs.~\ref{fig:efg_acfs}{\bf a}--{\bf c} feature the normalized  ACFs of  
$V^{\rm ext}_{\alpha\beta}$, $C_{\rm EFG}^{\rm norm}(t)$,
for Li$^+$, Na$^+$, K$^+$, Cl$^-$, Mg$^{2+}$ and Ca$^{2+}$
in the PIM, Madrid-2019 model, and the Amber14 FF parameters
for the SPC/E water and ions. Additional results for larger and 
more polarizable solutes such as Cs$^+$, Br$^-$, and I$^-$ as well as 
a comparison between the EFG ACFs for the alkali metals
are available in the Supporting~Fig.~S1. Interestingly,
the structure of the EFG ACFs remains similar across
the distinct FFs considered and is consistent
 with previous studies
\cite{RS1993, EngJon1981, EngJon1982, EngJon1984, Carof2014, Carof2015, Carof2016}. In general, $C_{\rm EFG}^{\rm norm}(t)$
relaxes in two steps with a pronounced short-range oscillatory regime for the smaller Li$^+$ and Mg$^{2+}$ ions, associated
with their tight confinement in the solvation shell \cite{Carof2015}. Furthermore, the ACF for Li$^+$ is positive
at all times in the PIM and Amber14 FF, whereas in the Madrid-2019
model it shows a negative region at around 0.1 ps.
In addition, a less pronounced oscillatory regime 
of $C_{\rm EFG}^{\rm norm}(t)$ at short times
is observed for the Ca$^{2+}$ ion as well. In the Amber14 FF,
as compared to the two other models,
the EFG ACF of Mg$^{2+}$ shows much more striking oscillations, whereas the decay for Ca$^{2+}$ is considerably slower.  This again
illustrates the difficulty for non-polarizable models with formal
ionic charges to accurately model multivalent ionic species\cite{DDijon2018, DDijon2020}. For larger ions, $C_{\rm EFG}^{\rm norm}(t)$
develops two clear relaxation steps. Interestingly, and consistently with AIMD simulations\cite{Philips2020}, $C_{\rm EFG}^{\rm norm}(t)$ develops a ``notch'' at very 
short times for the anions and for Cs$^+$, a feature associated with 
the librational motion of water molecules in the hydrogen bond
network\cite{Philips2020}. Finally, the reasonable agreement between 
$C_{\rm EFG}^{\rm norm}(t)$ obtained here and in AIMD\cite{Philips2017,Philips2020} confirms that the short time stretching and bending vibrational modes of water molecules (neglected in the rigid water models employed here) do not play a significant role in the form of the EFG relaxation.

\begin{figure}[t!]
\centering
\includegraphics[width=0.60\linewidth]{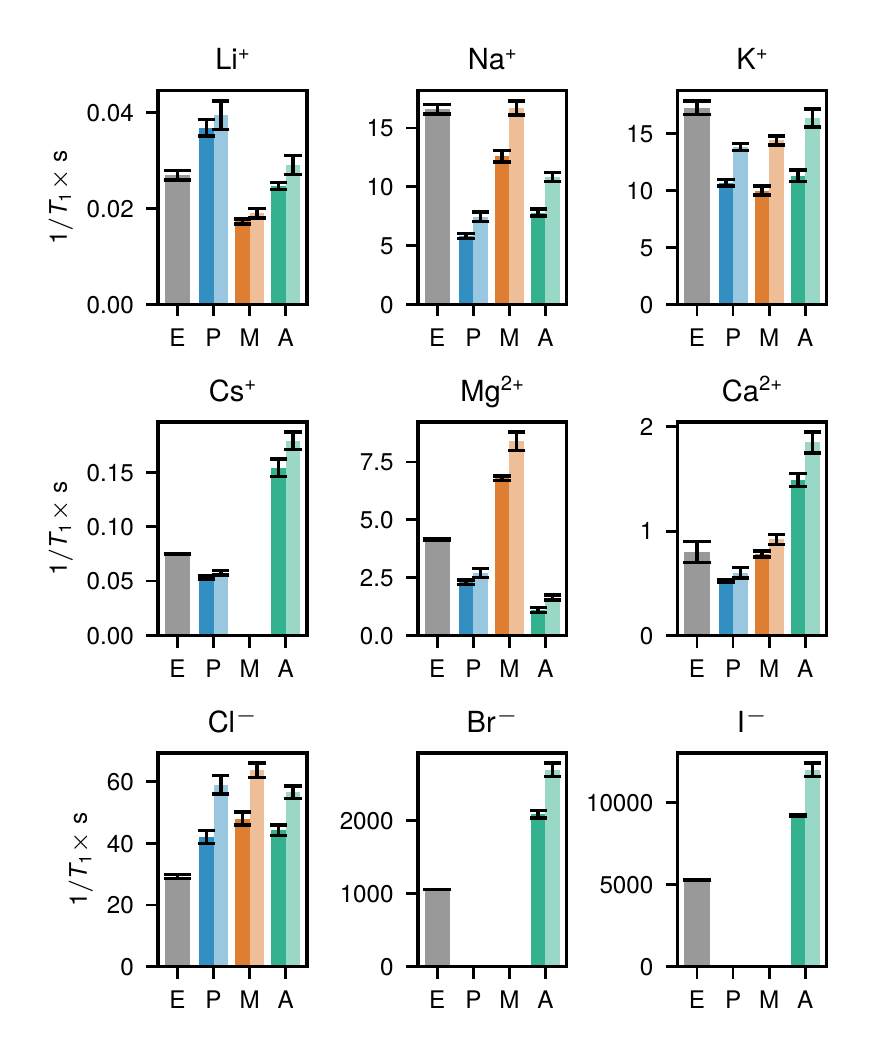}
\caption{\label{fig:nmr_rates}
Predicted NMR relaxation rates
of quadrupolar nuclei at infinite dilution. For every ion,
we indicate an experimental value (E), as well as the rate
obtained using the PIM (P),
the Madrid-2019 model (M),
and Amber14 FF parameters
for the SPC/E water and ions (A). For the P, M and E cases, the 
two adjacent columns indicate the $1/T_1$ value obtained using 
the EFG variance from the Sternheimer approximation $\langle \mathbf{V}_{\rm SA}^2 \rangle$
(left dark-colored columns) and directly from AI DFT
calculations $\langle \mathbf{V}_{\rm AI}^2 \rangle$
(right light-colored columns).
}
\end{figure}

As we show with running integrals of $C_{\rm EFG}^{\rm norm}(t)$
in Figs.~\ref{fig:efg_acfs}{\bf d}--{\bf f} and Supporting~Fig.~S1, 
accurate ACFs over around 10 ps (and even more in some cases)
are required to precisely measure the effective EFG correlation
time $\tau_c$ \eqref{eq:tauc}. All parameters of the EFG relaxation, in particular
$\langle \mathbf{V}_{\rm ext}^2 \rangle$ and $\tau_c$,  for all ions and 
FFs considered are again summarized in the Supporting~Tab.~S3.
Small differences in $\tau_c$
are found across the three FFs considered, yet in most cases
$\tau_c$ does not exceed 0.5 ps. For Ca$^{2+}$ in the Amber14 FF, 
$\tau_c = 1.06$ ps, likely being an overestimation in comparison to the two other FFs ($\tau_c = 0.31$ ps in the PIM 
and 0.44 ps in the Madrid-2019 for Ca$^{2+}$, see also Supporting~Fig.~S3). 

The variance of the external EFG $\langle \mathbf{V}_{\rm ext}^2 \rangle$ decreases horizontally along the periodic table, 
reflecting its dependence on the ionic size. Additionally, we list in the Supporting~Tab.~S3 the EFG variance 
$\langle \mathbf{V}_{\rm SA}^2 \rangle$ obtained with the Sternheimer approximation \eqref{eq:SA} using $\gamma_{\rm eff}$
from Tab.~\ref{tab:gamma_eff} as well as the variance
$\langle \mathbf{V}_{\rm AI}^2 \rangle$ obtained directly from 
the AI EFGs on our sets of 1000 configurations for each ion.
It is evident from this comparison that the failure of the Sternheimer approximation
in capturing the dispersion of the AI EFG (Fig.~\ref{fig:gsa}{\bf b}) always results in $\langle \mathbf{V}_{\rm SA}^2 \rangle$ 
being smaller than $\langle \mathbf{V}_{\rm AI}^2 \rangle$. In some 
cases, as shown further below, the use of $\langle \mathbf{V}_{\rm AI}^2 \rangle$ instead of $\langle \mathbf{V}_{\rm SA}^2 \rangle
\equiv (1 + \gamma_{\rm eff})^2 \langle \mathbf{V}_{\rm ext}^2 \rangle$ in Eq.~\eqref{eq:nmr_rates2} yields a better prediction
for the final NMR relaxation rates. 
 Nevertheless, a direct
replacement of $\langle \mathbf{V}_{\rm SA}^2 \rangle$ 
by $\langle \mathbf{V}_{\rm AI}^2 \rangle$ in Eq.~\eqref{eq:nmr_rates2} does not entirely account 
for the errors introduced by the Sternheimer approximation 
(see discussion in the SI).
 Consistently across the 
three FFs, a 25--35\% difference between $\langle \mathbf{V}_{\rm AI}^2 \rangle$ and $\langle \mathbf{V}_{\rm SA}^2 \rangle$  is
found for the Na$^+$, K$^+$, and Cl$^-$, potentially leading 
to a marked impact on the final relaxation rate.
In addition, a somewhat better correspondence between $\langle \mathbf{V}_{\rm SA}^2 \rangle$ and
$\langle \mathbf{V}_{\rm AI}^2 \rangle$ can be achieved if a direct fit between the squared EFGs
$\mathbf{V}_{\rm AI}^2 \simeq (1 + \gamma'_{\rm eff})^2 \mathbf{V}_{\rm ext}^2$ for $\gamma_{\rm eff}'$ is performed, generally resulting in larger values of  $\gamma'_{\rm eff}$ as compared to $\gamma_{\rm eff}$ in Tab.~\ref{tab:gamma_eff}. However, such a fit simultaneously yields bigger $\sigma(\mathbf{V}^2)$ and similarly fails to capture
the dispersion of the AI EFGs (not shown).
Finally, the differences in $\langle \mathbf{V}_{\rm ext}^2 \rangle$ and consequently in
$\langle \mathbf{V}_{\rm SA}^2 \rangle$ and $\langle \mathbf{V}_{\rm AI}^2 \rangle$ again highlight
the importance of the charge and density distributions around 
an ion on the resulting EFG at the ion position.

Finally, it is instructive
to compare the EFG relaxation parameters obtained here with classical MD with that from the AIMD studies\cite{Badu2013,Philips2017,Philips2020}. 
In Ref.~\citenum{Philips2017},  
$\tau_c = 0.26(0.06)$ ps and $\langle \mathbf{V}^2 \rangle = 0.21(0.01)$ (in a.u.) for $^{35}$Cl$^{-}$. In our case, for the same anion 
we find $\tau_c = 0.51(0.02)$ ps and 
$\langle \mathbf{V}_{\rm SA}^2 \rangle = 0.087(0.002)$
in the PIM; 
$\tau_c = 0.56(0.02)$ ps and 
$\langle \mathbf{V}_{\rm SA}^2 \rangle = 0.089(0.002)$
in the Madrid-2019 model; 
$\tau_c = 0.40(0.01)$ ps and 
$\langle \mathbf{V}_{\rm SA}^2 \rangle = 0.115(0.002)$
in the Amber14 FF. Similarly for $^{23}$Na$^{+}$, 
Ref.~\citenum{Philips2017} indicates $\tau_c = 0.13(0.03)$ ps and $\langle \mathbf{V}^2 \rangle = 0.028(0.001)$ (in a.u.), 
whereas in our case we obtain 
we find $\tau_c = 0.26(0.01)$ ps and 
$\langle \mathbf{V}_{\rm SA}^2 \rangle = 0.014(0.003)$
in the PIM; 
$\tau_c = 0.41(0.01)$ ps and 
$\langle \mathbf{V}_{\rm SA}^2 \rangle = 0.0197(0.0004)$
in the Madrid-2019 model; 
$\tau_c = 0.41(0.01)$ ps and 
$\langle \mathbf{V}_{\rm SA}^2 \rangle = 0.0122(0.0002)$
in the Amber14 FF. The AIMD results provide somewhat smaller estimates for $\tau_c$ when compared to classical MD, yet enhanced values of the EFG variance.
As shown below, classical MD can also yield quite good 
values for the NMR relaxation rates, provided that
$\gamma_{\rm eff}$, $\langle \mathbf{V}_{\rm ext}^2 \rangle$,
and $\tau_c$ are obtained consistently for a FF at hand.

\begin{table*}[t!]
\caption{\label{tab:nmr_rates}
Comparison between the NMR relaxation rates from experiments and classical MD simulations with various FFs 
for a series of quadrupolar ions at infinite dilution.}
\vspace*{2mm}
\small
\hspace*{-5mm}
\begin{tabular}{ccccccc}
\toprule
\multirow{2}{*}{Ion} & \multirow{2}{*}{$I$}
& \multirow{2}{*}{$Q$ (mb)\footnote{All values were taken from Ref.~\citenum{Pyykko2017}}} 
& \multicolumn{4}{c}{$1/T_1$ (s$^{-1}$)} \\ \cmidrule(lr){4-7}
 {} & {} & {}
 & Experiment & PIM 
 & Madrid-2019 &  Amber14 \\ \midrule
$^7$Li$^+$ & 3/2 & -40.1 & 
0.027\cite{HertzNaLi1974,Mohammadi2020} & 0.037(0.002) & 0.0173(0.0005) & 0.0247(0.0007)\\ \hline
$^{39}$K$^+$ & 3/2 & +60.3 & 16.7\cite{Fumino1997}, 17.8\cite{Weingaertner1977} & 10.7(0.3) & 10.0(0.4) & 11.3(0.5) \\ \hline
$^{23}$Na$^+$ & 3/2 & +104 & 17.0\cite{Fumino1997}, 16.2\cite{HertzNaLi1974} & 5.8(0.2) & 12.6(0.5) & 7.8(0.3) \\ \hline
$^{133}$Cs$^+$ & 7/2 & -3.43 & 0.075\cite{Hertz1973_1,Weingaertner1977} & 0.053(0.002) & --- & 0.154(0.008) \\ \hline
$^{35}$Cl$^-$  & 3/2 & -81.12 & 29.2(0.6)\cite{Struis1989}, 28\cite{Weingaertner1977} & 42.1(2.1) & 48.0(2.1) & 44.3(1.7) \\ \hline
$^{81}$Br$^-$  & 3/2 & 257.9 & 1050\cite{Hertz1973_1,Weingaertner1977}
 & --- & --- & 2080(50) \\ \hline
 $^{127}$I$^-$  & 5/2 & -688.22 & 5270\cite{Hertz1973_1}, 4600\cite{Weingaertner1977}
 & --- & --- & 9200(300) \\ \hline
$^{25}$Mg$^{2+}$ & 5/2 & +199.4 & 4.16(0.03)\cite{Struis1989} & 2.3(0.1) & 6.8(0.1) & 1.1(0.1) \\ \hline
$^{43}$Ca$^{2+}$ & 7/2 & -40.8 & 0.8(0.1)\cite{HelmHertz1981} & 0.52(0.01) & 0.78(0.03) & 1.49(0.06) \\ \bottomrule
\end{tabular}
\end{table*}

The final NMR relaxation rates $1/T_1$ for different ions 
obtained with the three classical FFs considered here are shown in Fig.~\ref{fig:nmr_rates} and also summarized in Tab.~\ref{tab:nmr_rates}, together
with the corresponding experimental values.
For $^7$Li$^+$, we find a quite good prediction for $1/T_1$ 
with the largest relative error 
$\left|1 - T_1^{\rm exp} / T_1^{\rm sim} \right|$
of around 35\% in both the PIM
and the Madrid-2019 model when compared to the experimental
value of 0.027 s$^{-1}$ that is measured for $^7$Li$^+$ in 
D$_2$O (which corresponds to the quadrupolar contribution only).
The value of 0.0247(0.0007) for $^7$Li$^+$ in the Amber14 FF
is in very good agreement with the experimental value. Recently,
the latter model has been used to determine the concentration
dependence of the quadrupolar contribution to the rate 
of $^7$Li$^+$ in simulations, yet a value of 
$\gamma_{\infty} = 0.17$ was used to account for the electronic cloud polarization effects and ultimately resulted in some 
differences between the simulation and experimental results 
at lower concentrations \cite{Mohammadi2020}. 
Here we show that this discrepancy can be overcome by considering
the model specific Sternheimer factor $\gamma_{\rm eff}$.

For $^{23}$Na$^+$, the Madrid-2019 model provides the best 
estimate for $1/T_1$ that is around 20\% smaller than the experimental value. For $^{39}$K$^+$, all FFs considered
yield quite close quadrupolar relaxation rates, being approximately 
40\% smaller than the experimental value. On the other hand, 
for the $^{35}$Cl$^-$ anion, the rates from all three FFs are
also quite close to each other but 40--60\% larger than those 
from the experiment. We find that the use of $\langle \mathbf{V}_{\rm AI}^2 \rangle$ instead of $\langle \mathbf{V}_{\rm SA}^2 \rangle$ improves the rate predictions for Na$^+$ and K$^+$,
yet leads to larger discrepancies for Cl$^-$ and the other two
anions Br$^-$ and I$^-$, the parameters for which are available in the Amber14 FF only. For the two divalent cations considered here, $^{25}$Mg$^{2+}$ and $^{43}$Ca$^{2+}$, we find that the PIM and
the Madrid-2019 model provide better estimates for the relaxation 
rates. In particular, very good agreement for with the experimental
rates is found $^{43}$Ca$^{2+}$ (35\% error in the PIM and 
almost quantitative agreement in the Madrid-2019 model), whereas 
a much higher discrepancy is observed in the Amber14 FF. Somewhat larger errors are found for $^{25}$Mg$^{2+}$,
however still better than in the Amber14 FF whose rate is 
around 4 times smaller than the experimental one. Finally, 
it is interesting to assess the rates obtained for larger and highly polarizable solutes like $^{133}$Cs$^+$, $^{81}$Br$^{-}$,
and $^{127}$I$^{-}$. The parameters for the latter three species
are available in the Amber14 FF, and in each case $1/T_1$ from 
the simulations is about twice as high as the experimental one.
On the contrary, for $^{133}$Cs$^+$ in the PIM, the resulting rate
$1/T_1^{\rm PIM} = 0.053(0.002)$ s$^{-1}$ is in reasonable agreement with the experimental one $1/T_1^{\rm exp} = 0.075$ s$^{-1}$, and is somewhat better than the state-of-the-art AIMD value obtained from the EFGs with relativistic effects included \cite{Philips2020}
$1/T_1^{\rm AIMD} = 0.033(0.006)$ s$^{-1}$.
This illustrates that classical MD is also suitable for computing the NMR relaxation
rates for divalent as well as large and strongly polarizable ionic species, provided that more
sophisticated methods to account for the electronic polarizability are employed.

\section{Discussion and concluding remarks}
\label{sec:disc}

We have shown in this work that accurate NMR relaxation
rates can be obtained from classical MD simulations, 
provided that local electron cloud polarization effects 
are taken into account consistently for each specific classical
FF considered. We have employed a Sternheimer-like parametrization for 
the electron cloud contribution to the EFG at the ion 
position by comparing the classical $V_{\alpha\beta}^{\rm ext}$ and quantum $V_{\alpha\beta}^{\rm AI}$ EFGs on a set 
of classically generated configurations.
We have found that a linear relationship between $V_{\alpha\beta}^{\rm AI}$
and $V_{\alpha\beta}^{\rm ext}$ holds well in all the cases
considered (Fig.~\ref{fig:sh_approx}), allowing to define 
effective model-dependent Sternheimer factors $\gamma_{\rm eff}$
(Fig.~\ref{fig:gamma_eff}). Yet, such Sternheimer factors
show a quite pronounced dependence on the FFs at hand. For instance, the 
difference in $\gamma_{\rm eff}$ across the 
classical models considered here can be up to 50\%, 
which is in some cases comparable to the difference between 
$\gamma_{\rm eff}$ and $\gamma_{\rm \infty}$ that was obtained 
for a more simplistic Watson sphere approximation.
We have found that considerable variations in $\gamma_{\rm eff}$ 
are due to the changes in the charge density representation of the employed water model
(that is, via the chosen set of point charges and dipoles),
whereas further differences are likely caused by variations in the 
solvation shell structure. In summary, $\gamma_{\rm eff}$ might 
be reasonably transferable for electrolyte FFs that employ the same water model, 
whereas the transferability is likely limited across FFs with different water models.

The error in the predicted
EFG variance using such an effective Sternheimer approximation \eqref{eq:SA} 
might be quite large (up to 50\% for certain ions) when compared to the AI EFGs (Tab.~\ref{tab:gamma_eff}). In particular,
the linear Sternheimer-like approximations that relate 
$V_{\alpha\beta}^{\rm AI}$ and $V_{\alpha\beta}^{\rm ext}$ 
fail to capture the large dispersion in $V_{\alpha\beta}^{\rm AI}$ (Fig.~\ref{fig:gsa}).
We have demonstrated that for the present
liquid matter systems it is very challenging to systematically reduce the prediction error, for instance by taking into account
non-linear effects \eqref{eq:efg_general2} that were shown to 
be essential for highly symmetric crystalline environments\cite{Calandra2002}. 
Some explanations can be proposed to rationalize this finding: $(i)$  
%in the present parametrization we employ $V_{\alpha\beta}^{\rm ext}$ generated by the classical %ionic charge distribution external to the given ion.
in the present parametrization we employ $V_{\alpha\beta}^{\rm ext}$ generated by the external
classical charge distribution around the given ion that can be very sensitive to the quality 
of the FF at hand -- while some FFs can provide a better approximation for the ab initio charge density, it is still hard to systematically assess its impact as the partitioning between internal and external charges includes some degree of arbitrariness in condensed phase systems;
$(ii)$ the local polarization effects
that have a considerable impact on the EFG at the nucleus might be strongly dependent on the instantaneous hydration shell structure
and solute-solvent interactions; $(iii)$ the inclusion of 
non-linear corrections to the EFG at the nucleus in liquid 
state systems can be limited by different symmetry properties 
of the $V_{\alpha\beta}$ and $E^{\rm ext}_{\alpha\beta}$ tensors.
In summary, it might be possible to construct an improved EFG parametrization
strategy by explicitly considering the local atomic environment around an ion, as this was done recently in the context of NMR for chemical shifts\cite{Chaker2019,Paruzzo2018}. In particular, correlating 
local solvent structure with the variable of interest (in this case,
the EFG tensor) using machine learning methods provides a possible direction to improve predictions from FF MD, as evidenced by recent
progress in the field of vibrational spectroscopy \cite{Kananenka2019,Baiz2021}.
The quality of the classical FF used for simulating the electrolyte
dynamics constitutes another important dimension of the problem. 
Here we have observed that the incorporation of electronic polarizability either through induced dipoles as in the PIM
or through scaled ionic charges as it the Madrid-2019 model
in many cases leads to better values of the NMR relaxation rates 
when compared to the non-polarizable Amber14 FF based on the SPC/E water and formal ionic charges. In particular, this is valid 
for the divalent Mg$^{2+}$ and Ca$^{2+}$ ions as well as for the large and highly polarizable Cs$^{+}$ ion (Fig.~\ref{fig:nmr_rates}). However, surprisingly, for smaller and less polarizable species, the non-polarizable Amber14 FF also provides quite good predictions for 
the final NMR relaxation rates. 
Furthermore, while the FFs used in this work employ rigid water molecules, 
the inclusion of stretching and bending modes in flexible water geometries might be necessary for a more accurate description of the
EFG relaxation at short time scales in classical MD. 
Quantitatively, the $1/T_1$ predictions obtained here are comparable 
and in some cases superior to the AIMD results\cite{Philips2017,Philips2020}. 
The fact that computationally
more efficient non-polarizable and scaled charge models are suitable 
for determining the NMR relaxation rates with reasonable 
accuracy, provided that a model-specific Sternheimer factor parametrized on ab initio calculations is used, paves the way for exploring the microscopic origins of many NMR relaxation phenomena, in particular those involving the concentration and pressure dependence of $1/T_1$ in concentrated electrolyte solutions and its relation to collective symmetry-breaking fluctuations in 
the solvation shell\cite{Carof2016}.

\section*{Associated content}

The Supporting Information is available free of charge at http://pubs.acs.org. \\

\noindent 
Additional discussion on the errors introduced by 
the Sternheimer approximation (Section S1);
EFG ACFs in an explicitly electroneutral system and sampled in the $NVE$ ensemble (Figure S1);
additional EFG ACFs and Sternheimer parametrization for
Cs$^+$, Br$^-$, and I$^-$ (Figure S2); 
comparison between the EFG ACFs and their integrals for Ca$^{2+}$ in 
different FFs (Figure S3);
the effect of swapping the charge density representation
between the PIM and Madrid-2019 FFs on the 
resulting $\gamma_{\rm eff}$ (Figures S4 and S5);
ion-oxygen RDFs (Figure S6); 
structure of the first solvation shell of ions 
in the three classical FFs considered (Figure S7);
coordination numbers of the first solvation shell for 
all ions in the three classical FFs considered
(Table S1); effective Sternheimer factors as a function of the number of water
molecules in the first solvation shell in different FFs (Table S2);
parameters of the EFG relaxation for all ions in the 
three classical FFs considered (Table S3).

\section*{Notes}
The authors declare no competing financial interest.

\section*{Acknowledgments}
The authors acknowledge fruitful discussions with Mathieu Salanne, Alexej Jerschow, Anne-Laure
Rollet, and Guillaume Mériguet. This project has received funding from the European Research Council (ERC)
under the European Union’s Horizon 2020 research and innovation programme (grant agreement No. 863473).

%\bibliography{biblio.bib}

\begin{mcitethebibliography}{65}
\providecommand*\natexlab[1]{#1}
\providecommand*\mciteSetBstSublistMode[1]{}
\providecommand*\mciteSetBstMaxWidthForm[2]{}
\providecommand*\mciteBstWouldAddEndPuncttrue
  {\def\EndOfBibitem{\unskip.}}
\providecommand*\mciteBstWouldAddEndPunctfalse
  {\let\EndOfBibitem\relax}
\providecommand*\mciteSetBstMidEndSepPunct[3]{}
\providecommand*\mciteSetBstSublistLabelBeginEnd[3]{}
\providecommand*\EndOfBibitem{}
\mciteSetBstSublistMode{f}
\mciteSetBstMaxWidthForm{subitem}{(\alph{mcitesubitemcount})}
\mciteSetBstSublistLabelBeginEnd
  {\mcitemaxwidthsubitemform\space}
  {\relax}
  {\relax}

\bibitem[Abragam(1961)]{abragam1961principles}
Abragam,~A. \emph{The Principles of Nuclear Magnetism}; Oxford university
  press, 1961\relax
\mciteBstWouldAddEndPuncttrue
\mciteSetBstMidEndSepPunct{\mcitedefaultmidpunct}
{\mcitedefaultendpunct}{\mcitedefaultseppunct}\relax
\EndOfBibitem
\bibitem[Roberts and Schnitker(1993)Roberts, and Schnitker]{RS1993}
Roberts,~J.~E.; Schnitker,~J. Ionic Quadrupolar Relaxation in Aqueous Solution:
  Dynamics of the Hydration Sphere. \emph{J. Phys. Chem.} \textbf{1993},
  \emph{97}, 5410--5417\relax
\mciteBstWouldAddEndPuncttrue
\mciteSetBstMidEndSepPunct{\mcitedefaultmidpunct}
{\mcitedefaultendpunct}{\mcitedefaultseppunct}\relax
\EndOfBibitem
\bibitem[Philips \latin{et~al.}(2017)Philips, Marchenko, Truflandier, and
  Autschbach]{Philips2017}
Philips,~A.; Marchenko,~A.; Truflandier,~L.~A.; Autschbach,~J. Quadrupolar
  {NMR} Relaxation from ab Initio Molecular Dynamics: Improved Sampling and
  Cluster Models versus Periodic Calculations. \emph{J. Chem. Theory Comput.}
  \textbf{2017}, \emph{13}, 4397--4409\relax
\mciteBstWouldAddEndPuncttrue
\mciteSetBstMidEndSepPunct{\mcitedefaultmidpunct}
{\mcitedefaultendpunct}{\mcitedefaultseppunct}\relax
\EndOfBibitem
\bibitem[Philips \latin{et~al.}(2019)Philips, Marchenko, Ducati, and
  Autschbach]{Philips2018}
Philips,~A.; Marchenko,~A.; Ducati,~L.~C.; Autschbach,~J. Quadrupolar $^{14}$N
  NMR Relaxation from Force-Field and Ab Initio Molecular Dynamics in Different
  Solvents. \emph{J. Chem. Theory Comput.} \textbf{2019}, \emph{15},
  509--519\relax
\mciteBstWouldAddEndPuncttrue
\mciteSetBstMidEndSepPunct{\mcitedefaultmidpunct}
{\mcitedefaultendpunct}{\mcitedefaultseppunct}\relax
\EndOfBibitem
\bibitem[Philips and Autschbach(2020)Philips, and Autschbach]{Philips2020}
Philips,~A.; Autschbach,~J. Quadrupolar NMR Relaxation of Aqueous
  $^{127}$I$^-$, $^{131}$Xe$^+$, and $^{133}$Cs$^+$: A First-Principles
  Approach from Dynamics to Properties. \emph{J. Chem. Theory Comput.}
  \textbf{2020}, \emph{16}, 5835--5844\relax
\mciteBstWouldAddEndPuncttrue
\mciteSetBstMidEndSepPunct{\mcitedefaultmidpunct}
{\mcitedefaultendpunct}{\mcitedefaultseppunct}\relax
\EndOfBibitem
\bibitem[Badu \latin{et~al.}(2013)Badu, Truflandier, and Autschbach]{Badu2013}
Badu,~S.; Truflandier,~L.; Autschbach,~J. Quadrupolar NMR Spin Relaxation
  Calculated Using Ab Initio Molecular Dynamics: Group 1 and Group 17 Ions in
  Aqueous Solution. \emph{J. Chem. Theory Comput.} \textbf{2013}, \emph{9},
  4074--4086\relax
\mciteBstWouldAddEndPuncttrue
\mciteSetBstMidEndSepPunct{\mcitedefaultmidpunct}
{\mcitedefaultendpunct}{\mcitedefaultseppunct}\relax
\EndOfBibitem
\bibitem[Engström and Jönsson(1981)Engström, and Jönsson]{EngJon1981}
Engström,~S.; Jönsson,~B. {Monte Carlo} Simulations of the Electric Field
  Gradient Fluctuation at the Nucleus of a Lithium Ion in Dilute Aqueous
  Solution. \emph{Mol. Phys.} \textbf{1981}, \emph{43}, 1235--1253\relax
\mciteBstWouldAddEndPuncttrue
\mciteSetBstMidEndSepPunct{\mcitedefaultmidpunct}
{\mcitedefaultendpunct}{\mcitedefaultseppunct}\relax
\EndOfBibitem
\bibitem[Engström \latin{et~al.}(1982)Engström, Jönsson, and
  Jönsson]{EngJon1982}
Engström,~S.; Jönsson,~B.; Jönsson,~B. A Molecular Approach to Quadrupole
  Relaxation. {Monte Carlo} Simulations of Dilute {Li}$^+$, {Na}$^+$, and
  {Cl}$^-$ Aqueous Solutions. \emph{J. Magn. Reson. (1969-1992)} \textbf{1982},
  \emph{50}, 1--20\relax
\mciteBstWouldAddEndPuncttrue
\mciteSetBstMidEndSepPunct{\mcitedefaultmidpunct}
{\mcitedefaultendpunct}{\mcitedefaultseppunct}\relax
\EndOfBibitem
\bibitem[Engström \latin{et~al.}(1984)Engström, Jönsson, and
  Impey]{EngJon1984}
Engström,~S.; Jönsson,~B.; Impey,~R.~W. Molecular Dynamic Simulation of
  Quadrupole Relaxation of Atomic Ions in Aqueous Solution. \emph{J. Chem.
  Phys.} \textbf{1984}, \emph{80}, 5481--5486\relax
\mciteBstWouldAddEndPuncttrue
\mciteSetBstMidEndSepPunct{\mcitedefaultmidpunct}
{\mcitedefaultendpunct}{\mcitedefaultseppunct}\relax
\EndOfBibitem
\bibitem[Carof \latin{et~al.}(2014)Carof, Salanne, Charpentier, and
  Rotenberg]{Carof2014}
Carof,~A.; Salanne,~M.; Charpentier,~T.; Rotenberg,~B. Accurate Quadrupolar
  {NMR} Relaxation Rates of Aqueous Cations from Classical Molecular Dynamics.
  \emph{J. Phys. Chem. B} \textbf{2014}, \emph{118}, 13252--13257\relax
\mciteBstWouldAddEndPuncttrue
\mciteSetBstMidEndSepPunct{\mcitedefaultmidpunct}
{\mcitedefaultendpunct}{\mcitedefaultseppunct}\relax
\EndOfBibitem
\bibitem[Mohammadi \latin{et~al.}(2020)Mohammadi, Benders, and
  Jerschow]{Mohammadi2020}
Mohammadi,~M.; Benders,~S.; Jerschow,~A. Nuclear Magnetic Resonance
  Spin-Lattice Relaxation of Lithium Ions in Aqueous Solution by {NMR} and
  Molecular Dynamics. \emph{J. Chem. Phys.} \textbf{2020}, \emph{153},
  184502\relax
\mciteBstWouldAddEndPuncttrue
\mciteSetBstMidEndSepPunct{\mcitedefaultmidpunct}
{\mcitedefaultendpunct}{\mcitedefaultseppunct}\relax
\EndOfBibitem
\bibitem[Autschbach \latin{et~al.}(2010)Autschbach, Zheng, and
  Schurko]{Autschbach2010}
Autschbach,~J.; Zheng,~S.; Schurko,~R.~W. Analysis of electric field gradient
  tensors at quadrupolar nuclei in common structural motifs. \emph{Concepts
  Magn. Reson. A} \textbf{2010}, \emph{36A}, 84--126\relax
\mciteBstWouldAddEndPuncttrue
\mciteSetBstMidEndSepPunct{\mcitedefaultmidpunct}
{\mcitedefaultendpunct}{\mcitedefaultseppunct}\relax
\EndOfBibitem
\bibitem[Hertz(1973)]{Hertz1973_1}
Hertz,~H.~G. Magnetic Relaxation by Quadrupole Interaction of Ionic Nuclei in
  Electrolyte Solutions Part I: Limiting Values for Infinite Dilution.
  \emph{Ber. Bunsenges. Phys. Chem.} \textbf{1973}, \emph{77}, 531--540\relax
\mciteBstWouldAddEndPuncttrue
\mciteSetBstMidEndSepPunct{\mcitedefaultmidpunct}
{\mcitedefaultendpunct}{\mcitedefaultseppunct}\relax
\EndOfBibitem
\bibitem[Hertz(1973)]{Hertz1973_2}
Hertz,~H.~G. Magnetic Relaxation by Quadrupole Interaction of Ionic Nuclei in
  Electrolyte Solutions Part II: Relaxation at Finite Ion Concentrations.
  \emph{Ber. Bunsenges. Phys. Chem.} \textbf{1973}, \emph{77}, 688--697\relax
\mciteBstWouldAddEndPuncttrue
\mciteSetBstMidEndSepPunct{\mcitedefaultmidpunct}
{\mcitedefaultendpunct}{\mcitedefaultseppunct}\relax
\EndOfBibitem
\bibitem[Valiev and Khabibullin(1961)Valiev, and Khabibullin]{Valiev1961}
Valiev,~K.; Khabibullin,~B. The Nuclear Magnetic Resonance and the Structure of
  Aqueous Electrolyte Solutions. \emph{Russ. J. Phys. Chem.} \textbf{1961},
  \emph{35}, 2265--2274\relax
\mciteBstWouldAddEndPuncttrue
\mciteSetBstMidEndSepPunct{\mcitedefaultmidpunct}
{\mcitedefaultendpunct}{\mcitedefaultseppunct}\relax
\EndOfBibitem
\bibitem[Friedman(1978)]{Friedman}
Friedman,~H. The 2 Picosecond Motion in the Hydration Shell of {Ni}$^{2+}$.
  Protons and Ions Involved in Fast Dynamic Phenomena. 1978\relax
\mciteBstWouldAddEndPuncttrue
\mciteSetBstMidEndSepPunct{\mcitedefaultmidpunct}
{\mcitedefaultendpunct}{\mcitedefaultseppunct}\relax
\EndOfBibitem
\bibitem[Sternheimer(1950)]{Sternheimer1950}
Sternheimer,~R. On Nuclear Quadrupole Moments. \emph{Phys. Rev.} \textbf{1950},
  \emph{80}, 102--103\relax
\mciteBstWouldAddEndPuncttrue
\mciteSetBstMidEndSepPunct{\mcitedefaultmidpunct}
{\mcitedefaultendpunct}{\mcitedefaultseppunct}\relax
\EndOfBibitem
\bibitem[Foley \latin{et~al.}(1954)Foley, Sternheimer, and Tycko]{Foley1954}
Foley,~H.~M.; Sternheimer,~R.~M.; Tycko,~D. Nuclear Quadrupole Coupling in
  Polar Molecules. \emph{Phys. Rev.} \textbf{1954}, \emph{93}, 734--742\relax
\mciteBstWouldAddEndPuncttrue
\mciteSetBstMidEndSepPunct{\mcitedefaultmidpunct}
{\mcitedefaultendpunct}{\mcitedefaultseppunct}\relax
\EndOfBibitem
\bibitem[Sternheimer(1966)]{Sternheimer1966}
Sternheimer,~R.~M. Shielding and Antishielding Effects for Various Ions and
  Atomic Systems. \emph{Phys. Rev.} \textbf{1966}, \emph{146}, 140--160\relax
\mciteBstWouldAddEndPuncttrue
\mciteSetBstMidEndSepPunct{\mcitedefaultmidpunct}
{\mcitedefaultendpunct}{\mcitedefaultseppunct}\relax
\EndOfBibitem
\bibitem[Schmidt \latin{et~al.}(1980)Schmidt, Sen, Das, and Weiss]{Schmidt1980}
Schmidt,~P.~C.; Sen,~K.~D.; Das,~T.~P.; Weiss,~A. Effect of Self-Consistency
  and Crystalline Potential in the Solid State on Nuclear Quadrupole
  Sternheimer Antishielding Factors in Closed-Shell Ions. \emph{Phys. Rev. B}
  \textbf{1980}, \emph{22}, 4167--4179\relax
\mciteBstWouldAddEndPuncttrue
\mciteSetBstMidEndSepPunct{\mcitedefaultmidpunct}
{\mcitedefaultendpunct}{\mcitedefaultseppunct}\relax
\EndOfBibitem
\bibitem[Tazi \latin{et~al.}(2012)Tazi, Molina, Rotenberg, Turq, Vuilleumier,
  and Salanne]{PIM}
Tazi,~S.; Molina,~J.~J.; Rotenberg,~B.; Turq,~P.; Vuilleumier,~R.; Salanne,~M.
  A Transferable \emph{Ab Initio} Based Force Field for Aqueous Ions. \emph{J.
  Chem. Phys.} \textbf{2012}, \emph{136}, 114507\relax
\mciteBstWouldAddEndPuncttrue
\mciteSetBstMidEndSepPunct{\mcitedefaultmidpunct}
{\mcitedefaultendpunct}{\mcitedefaultseppunct}\relax
\EndOfBibitem
\bibitem[Dang and Chang(1997)Dang, and Chang]{DangChang}
Dang,~L.~X.; Chang,~T.-M. Molecular Dynamics Study of Water Clusters, Liquid,
  and Liquid–Vapor Interface of Water With Many-Body Potentials. \emph{J.
  Chem. Phys.} \textbf{1997}, \emph{106}, 8149--8159\relax
\mciteBstWouldAddEndPuncttrue
\mciteSetBstMidEndSepPunct{\mcitedefaultmidpunct}
{\mcitedefaultendpunct}{\mcitedefaultseppunct}\relax
\EndOfBibitem
\bibitem[Zeron \latin{et~al.}(2019)Zeron, Abascal, and Vega]{Madrid_2019}
Zeron,~I.~M.; Abascal,~J. L.~F.; Vega,~C. {A Force Field of Li$^+$, Na$^+$,
  K$^+$, Mg$^{2+}$, Ca$^{2+}$, Cl$^-$, and SO$_4^{2-}$ in Aqueous Solution
  Based on the TIP4P/2005 Water Model and Scaled Charges for the Ions}.
  \emph{J. Chem. Phys.} \textbf{2019}, \emph{151}, 134504\relax
\mciteBstWouldAddEndPuncttrue
\mciteSetBstMidEndSepPunct{\mcitedefaultmidpunct}
{\mcitedefaultendpunct}{\mcitedefaultseppunct}\relax
\EndOfBibitem
\bibitem[Abascal and Vega(2005)Abascal, and Vega]{TIP4P2005}
Abascal,~J. L.~F.; Vega,~C. A General Purpose Model for the Condensed Phases of
  Water: {TIP4P/2005}. \emph{J. Chem. Phys.} \textbf{2005}, \emph{123},
  234505\relax
\mciteBstWouldAddEndPuncttrue
\mciteSetBstMidEndSepPunct{\mcitedefaultmidpunct}
{\mcitedefaultendpunct}{\mcitedefaultseppunct}\relax
\EndOfBibitem
\bibitem[Joung and Cheatham(2008)Joung, and Cheatham]{Joung2008}
Joung,~I.~S.; Cheatham,~T.~E. Determination of Alkali and Halide Monovalent Ion
  Parameters for Use in Explicitly Solvated Biomolecular Simulations. \emph{J.
  Phys. Chem. B} \textbf{2008}, \emph{112}, 9020--9041\relax
\mciteBstWouldAddEndPuncttrue
\mciteSetBstMidEndSepPunct{\mcitedefaultmidpunct}
{\mcitedefaultendpunct}{\mcitedefaultseppunct}\relax
\EndOfBibitem
\bibitem[Joung and Cheatham(2009)Joung, and Cheatham]{Joung2009}
Joung,~I.~S.; Cheatham,~T.~E. Molecular Dynamics Simulations of the Dynamic and
  Energetic Properties of Alkali and Halide Ions Using Water-Model-Specific Ion
  Parameters. \emph{J. Phys. Chem. B} \textbf{2009}, \emph{113},
  13279--13290\relax
\mciteBstWouldAddEndPuncttrue
\mciteSetBstMidEndSepPunct{\mcitedefaultmidpunct}
{\mcitedefaultendpunct}{\mcitedefaultseppunct}\relax
\EndOfBibitem
\bibitem[Li \latin{et~al.}(2013)Li, Roberts, Chakravorty, and Merz]{Li2013}
Li,~P.; Roberts,~B.~P.; Chakravorty,~D.~K.; Merz,~K.~M. Rational Design of
  Particle Mesh Ewald Compatible {Lennard-Jones} Parameters for +2 Metal
  Cations in Explicit Solvent. \emph{J. Chem. Theor. Comput.} \textbf{2013},
  \emph{9}, 2733--2748\relax
\mciteBstWouldAddEndPuncttrue
\mciteSetBstMidEndSepPunct{\mcitedefaultmidpunct}
{\mcitedefaultendpunct}{\mcitedefaultseppunct}\relax
\EndOfBibitem
\bibitem[Maier \latin{et~al.}(2015)Maier, Martinez, Kasavajhala, Wickstrom,
  Hauser, and Simmerling]{AMBER14}
Maier,~J.~A.; Martinez,~C.; Kasavajhala,~K.; Wickstrom,~L.; Hauser,~K.~E.;
  Simmerling,~C. {ff14SB}: Improving the Accuracy of Protein Side Chain and
  Backbone Parameters from {ff99SB}. \emph{J. Chem. Theory Comput.}
  \textbf{2015}, \emph{11}, 3696--3713\relax
\mciteBstWouldAddEndPuncttrue
\mciteSetBstMidEndSepPunct{\mcitedefaultmidpunct}
{\mcitedefaultendpunct}{\mcitedefaultseppunct}\relax
\EndOfBibitem
\bibitem[Berendsen \latin{et~al.}(1987)Berendsen, Grigera, and Straatsma]{SPCE}
Berendsen,~H. J.~C.; Grigera,~J.~R.; Straatsma,~T.~P. The Missing Term in
  Effective Pair Potentials. \emph{J. Phys. Chem.} \textbf{1987}, \emph{91},
  6269--6271\relax
\mciteBstWouldAddEndPuncttrue
\mciteSetBstMidEndSepPunct{\mcitedefaultmidpunct}
{\mcitedefaultendpunct}{\mcitedefaultseppunct}\relax
\EndOfBibitem
\bibitem[Nosé(1984)]{Nose1984}
Nosé,~S. A Unified Formulation of the Constant Temperature Molecular Dynamics
  Methods. \emph{J. Chem. Phys.} \textbf{1984}, \emph{81}, 511--519\relax
\mciteBstWouldAddEndPuncttrue
\mciteSetBstMidEndSepPunct{\mcitedefaultmidpunct}
{\mcitedefaultendpunct}{\mcitedefaultseppunct}\relax
\EndOfBibitem
\bibitem[Hoover(1985)]{Hoover1985}
Hoover,~W.~G. Canonical Dynamics: Equilibrium Phase-Space Distributions.
  \emph{Phys. Rev. A} \textbf{1985}, \emph{31}, 1695--1697\relax
\mciteBstWouldAddEndPuncttrue
\mciteSetBstMidEndSepPunct{\mcitedefaultmidpunct}
{\mcitedefaultendpunct}{\mcitedefaultseppunct}\relax
\EndOfBibitem
\bibitem[Martyna \latin{et~al.}(1992)Martyna, Klein, and Tuckerman]{NHchains}
Martyna,~G.~J.; Klein,~M.~L.; Tuckerman,~M. Nosé–Hoover Chains: The
  Canonical Ensemble via Continuous Dynamics. \emph{J. Chem. Phys.}
  \textbf{1992}, \emph{97}, 2635--2643\relax
\mciteBstWouldAddEndPuncttrue
\mciteSetBstMidEndSepPunct{\mcitedefaultmidpunct}
{\mcitedefaultendpunct}{\mcitedefaultseppunct}\relax
\EndOfBibitem
\bibitem[Aguado and Madden(2003)Aguado, and Madden]{AguadoMadden}
Aguado,~A.; Madden,~P.~A. Ewald Summation of Electrostatic Multipole
  Interactions up to the Quadrupolar Level. \emph{J. Chem. Phys} \textbf{2003},
  \emph{119}, 7471--7483\relax
\mciteBstWouldAddEndPuncttrue
\mciteSetBstMidEndSepPunct{\mcitedefaultmidpunct}
{\mcitedefaultendpunct}{\mcitedefaultseppunct}\relax
\EndOfBibitem
\bibitem[Laino and Hutter(2008)Laino, and Hutter]{Laino2003}
Laino,~T.; Hutter,~J. Notes on {“Ewald Summation of Electrostatic Multipole
  Interactions up to Quadrupolar Level” [J. Chem. Phys. 119, 7471 (2003)]}.
  \emph{J. Chem. Phys.} \textbf{2008}, \emph{129}, 074102\relax
\mciteBstWouldAddEndPuncttrue
\mciteSetBstMidEndSepPunct{\mcitedefaultmidpunct}
{\mcitedefaultendpunct}{\mcitedefaultseppunct}\relax
\EndOfBibitem
\bibitem[Andersen(1983)]{Rattle}
Andersen,~H.~C. Rattle: A “Velocity” Version of the Shake Algorithm for
  Molecular Dynamics Calculations. \emph{J. Comput. Phys.} \textbf{1983},
  \emph{52}, 24--34\relax
\mciteBstWouldAddEndPuncttrue
\mciteSetBstMidEndSepPunct{\mcitedefaultmidpunct}
{\mcitedefaultendpunct}{\mcitedefaultseppunct}\relax
\EndOfBibitem
\bibitem[Marin-Laflèche \latin{et~al.}(2020)Marin-Laflèche, Haefele, Scalfi,
  Coretti, Dufils, Jeanmairet, Reed, Serva, Berthin, Bacon, Bonella, Rotenberg,
  Madden, and Salanne]{MW2}
Marin-Laflèche,~A.; Haefele,~M.; Scalfi,~L.; Coretti,~A.; Dufils,~T.;
  Jeanmairet,~G.; Reed,~S.~K.; Serva,~A.; Berthin,~R.; Bacon,~C.; Bonella,~S.;
  Rotenberg,~B.; Madden,~P.~A.; Salanne,~M. MetalWalls: A Classical Molecular
  Dynamics Software Dedicated to the Simulation of Electrochemical Systems.
  \emph{J. Open Source Softw.} \textbf{2020}, \emph{5}, 2373\relax
\mciteBstWouldAddEndPuncttrue
\mciteSetBstMidEndSepPunct{\mcitedefaultmidpunct}
{\mcitedefaultendpunct}{\mcitedefaultseppunct}\relax
\EndOfBibitem
\bibitem[Marin-Laflèche \latin{et~al.}(2021)Marin-Laflèche, Haefele, Scalfi,
  Coretti, Chubak, Dufils, Jeanmairet, Reed, Serva, Berthin, Bacon, Bonella,
  Rotenberg, Madden, and Salanne]{MW_2106}
Marin-Laflèche,~A.; Haefele,~M.; Scalfi,~L.; Coretti,~A.; Chubak,~I.;
  Dufils,~T.; Jeanmairet,~G.; Reed,~S.; Serva,~A.; Berthin,~R.; Bacon,~C.;
  Bonella,~S.; Rotenberg,~B.; Madden,~P.; Salanne,~M. MetalWalls 21.06. 2021;
  \url{https://doi.org/10.5281/zenodo.4912611}\relax
\mciteBstWouldAddEndPuncttrue
\mciteSetBstMidEndSepPunct{\mcitedefaultmidpunct}
{\mcitedefaultendpunct}{\mcitedefaultseppunct}\relax
\EndOfBibitem
\bibitem[Giannozzi \latin{et~al.}(2009)Giannozzi, Baroni, Bonini, Calandra,
  Car, Cavazzoni, Ceresoli, Chiarotti, Cococcioni, Dabo, Corso, de~Gironcoli,
  Fabris, Fratesi, Gebauer, Gerstmann, Gougoussis, Kokalj, Lazzeri,
  Martin-Samos, Marzari, Mauri, Mazzarello, Paolini, Pasquarello, Paulatto,
  Sbraccia, Scandolo, Sclauzero, Seitsonen, Smogunov, Umari, and
  Wentzcovitch]{Giannozzi_2009}
Giannozzi,~P.; Baroni,~S.; Bonini,~N.; Calandra,~M.; Car,~R.; Cavazzoni,~C.;
  Ceresoli,~D.; Chiarotti,~G.~L.; Cococcioni,~M.; Dabo,~I.; Corso,~A.~D.;
  de~Gironcoli,~S.; Fabris,~S.; Fratesi,~G.; Gebauer,~R.; Gerstmann,~U.;
  Gougoussis,~C.; Kokalj,~A.; Lazzeri,~M.; Martin-Samos,~L.; Marzari,~N.;
  Mauri,~F.; Mazzarello,~R.; Paolini,~S.; Pasquarello,~A.; Paulatto,~L.;
  Sbraccia,~C.; Scandolo,~S.; Sclauzero,~G.; Seitsonen,~A.~P.; Smogunov,~A.;
  Umari,~P.; Wentzcovitch,~R.~M. {QUANTUM} {ESPRESSO}: A Modular and
  Open-Source Software Project for Quantum Simulations of Materials. \emph{J.
  Phys. Condens. Matter} \textbf{2009}, \emph{21}, 395502\relax
\mciteBstWouldAddEndPuncttrue
\mciteSetBstMidEndSepPunct{\mcitedefaultmidpunct}
{\mcitedefaultendpunct}{\mcitedefaultseppunct}\relax
\EndOfBibitem
\bibitem[Bl\"ochl(1994)]{Bloechl1994}
Bl\"ochl,~P.~E. Projector Augmented-Wave Method. \emph{Phys. Rev. B}
  \textbf{1994}, \emph{50}, 17953--17979\relax
\mciteBstWouldAddEndPuncttrue
\mciteSetBstMidEndSepPunct{\mcitedefaultmidpunct}
{\mcitedefaultendpunct}{\mcitedefaultseppunct}\relax
\EndOfBibitem
\bibitem[Petrilli \latin{et~al.}(1998)Petrilli, Bl\"ochl, Blaha, and
  Schwarz]{Petrilli1998}
Petrilli,~H.~M.; Bl\"ochl,~P.~E.; Blaha,~P.; Schwarz,~K.
  Electric-field-gradient calculations using the projector augmented wave
  method. \emph{Phys. Rev. B} \textbf{1998}, \emph{57}, 14690--14697\relax
\mciteBstWouldAddEndPuncttrue
\mciteSetBstMidEndSepPunct{\mcitedefaultmidpunct}
{\mcitedefaultendpunct}{\mcitedefaultseppunct}\relax
\EndOfBibitem
\bibitem[Charpentier(2011)]{Charpentier2011}
Charpentier,~T. The {PAW/GIPAW} Approach for Computing {NMR} Parameters: A New
  Dimension Added to {NMR} Study of Solids. \emph{Solid State Nucl. Magn.
  Reson.} \textbf{2011}, \emph{40}, 1--20\relax
\mciteBstWouldAddEndPuncttrue
\mciteSetBstMidEndSepPunct{\mcitedefaultmidpunct}
{\mcitedefaultendpunct}{\mcitedefaultseppunct}\relax
\EndOfBibitem
\bibitem[Varini \latin{et~al.}(2013)Varini, Ceresoli, Martin-Samos, Girotto,
  and Cavazzoni]{Varini2013}
Varini,~N.; Ceresoli,~D.; Martin-Samos,~L.; Girotto,~I.; Cavazzoni,~C.
  Enhancement of DFT-calculations at petascale: Nuclear Magnetic Resonance,
  Hybrid Density Functional Theory and Car–Parrinello calculations.
  \emph{Comput. Phys. Commun.} \textbf{2013}, \emph{184}, 1827--1833\relax
\mciteBstWouldAddEndPuncttrue
\mciteSetBstMidEndSepPunct{\mcitedefaultmidpunct}
{\mcitedefaultendpunct}{\mcitedefaultseppunct}\relax
\EndOfBibitem
\bibitem[Perdew \latin{et~al.}(1996)Perdew, Burke, and Ernzerhof]{PBE1996}
Perdew,~J.~P.; Burke,~K.; Ernzerhof,~M. Generalized Gradient Approximation Made
  Simple. \emph{Phys. Rev. Lett.} \textbf{1996}, \emph{77}, 3865--3868\relax
\mciteBstWouldAddEndPuncttrue
\mciteSetBstMidEndSepPunct{\mcitedefaultmidpunct}
{\mcitedefaultendpunct}{\mcitedefaultseppunct}\relax
\EndOfBibitem
\bibitem[GIP()]{GIPAW_pseudo}
{GIPAW} Norm-Conserving Pseudopotentials. \url{https://sites.
  google.com/site/dceresoli/pseudopotentials}, Accessed: 2020-04-28\relax
\mciteBstWouldAddEndPuncttrue
\mciteSetBstMidEndSepPunct{\mcitedefaultmidpunct}
{\mcitedefaultendpunct}{\mcitedefaultseppunct}\relax
\EndOfBibitem
\bibitem[{Dal Corso}(2014)]{pslibrary}
{Dal Corso},~A. Pseudopotentials Periodic Table: From {H} to {Pu}.
  \emph{Comput. Mater. Sci.} \textbf{2014}, \emph{95}, 337--350\relax
\mciteBstWouldAddEndPuncttrue
\mciteSetBstMidEndSepPunct{\mcitedefaultmidpunct}
{\mcitedefaultendpunct}{\mcitedefaultseppunct}\relax
\EndOfBibitem
\bibitem[Kathmann \latin{et~al.}(2011)Kathmann, Kuo, Mundy, and
  Schenter]{Kathmann2011}
Kathmann,~S.~M.; Kuo,~I.-F.~W.; Mundy,~C.~J.; Schenter,~G.~K. Understanding the
  Surface Potential of Water. \emph{J. Phys. Chem. B} \textbf{2011},
  \emph{115}, 4369--4377\relax
\mciteBstWouldAddEndPuncttrue
\mciteSetBstMidEndSepPunct{\mcitedefaultmidpunct}
{\mcitedefaultendpunct}{\mcitedefaultseppunct}\relax
\EndOfBibitem
\bibitem[Carof \latin{et~al.}(2015)Carof, Salanne, Charpentier, and
  Rotenberg]{Carof2015}
Carof,~A.; Salanne,~M.; Charpentier,~T.; Rotenberg,~B. On the Microscopic
  Fluctuations Driving the NMR Relaxation of Quadrupolar Ions in Water.
  \emph{J. Chem. Phys.} \textbf{2015}, \emph{143}, 194504\relax
\mciteBstWouldAddEndPuncttrue
\mciteSetBstMidEndSepPunct{\mcitedefaultmidpunct}
{\mcitedefaultendpunct}{\mcitedefaultseppunct}\relax
\EndOfBibitem
\bibitem[Fowler \latin{et~al.}(1989)Fowler, Lazzeretti, Steiner, and
  Zanasi]{Fowler1989}
Fowler,~P.; Lazzeretti,~P.; Steiner,~E.; Zanasi,~R. The Theory of Sternheimer
  Shielding in Molecules in External Fields. \emph{Chem. Phys.} \textbf{1989},
  \emph{133}, 221--235\relax
\mciteBstWouldAddEndPuncttrue
\mciteSetBstMidEndSepPunct{\mcitedefaultmidpunct}
{\mcitedefaultendpunct}{\mcitedefaultseppunct}\relax
\EndOfBibitem
\bibitem[Calandra \latin{et~al.}(2002)Calandra, Domene, Fowler, and
  Madden]{Calandra2002}
Calandra,~P.; Domene,~C.; Fowler,~P.~W.; Madden,~P.~A. Nuclear Quadrupole
  Coupling of $^17$O and $^33$S in Ionic Solids: Invalidation of the
  Sternheimer Model by Short-Range Corrections. \emph{J. Phys. Chem. B}
  \textbf{2002}, \emph{106}, 10342--10348\relax
\mciteBstWouldAddEndPuncttrue
\mciteSetBstMidEndSepPunct{\mcitedefaultmidpunct}
{\mcitedefaultendpunct}{\mcitedefaultseppunct}\relax
\EndOfBibitem
\bibitem[Buckingham(1967)]{Buckingham1967}
Buckingham,~A.~D. \emph{Advances in Chemical Physics}; John Wiley \& Sons, Ltd,
  1967; pp 107--142\relax
\mciteBstWouldAddEndPuncttrue
\mciteSetBstMidEndSepPunct{\mcitedefaultmidpunct}
{\mcitedefaultendpunct}{\mcitedefaultseppunct}\relax
\EndOfBibitem
\bibitem[Carof(2015)]{CarofPhd}
Carof,~A. Modélisation de la Relaxométrie RMN pour des Ions Mono-Atomiques
  Quadrupolaires en Phase Condensée. Ph.D.\ thesis, Université Pierre et
  Marie Curie, Paris, 2015\relax
\mciteBstWouldAddEndPuncttrue
\mciteSetBstMidEndSepPunct{\mcitedefaultmidpunct}
{\mcitedefaultendpunct}{\mcitedefaultseppunct}\relax
\EndOfBibitem
\bibitem[Carof \latin{et~al.}(2016)Carof, Salanne, Charpentier, and
  Rotenberg]{Carof2016}
Carof,~A.; Salanne,~M.; Charpentier,~T.; Rotenberg,~B. Collective Water
  Dynamics in the First Solvation Shell Drive the {NMR} Relaxation of Aqueous
  Quadrupolar Cations. \emph{J. Chem. Phys.} \textbf{2016}, \emph{145},
  124508\relax
\mciteBstWouldAddEndPuncttrue
\mciteSetBstMidEndSepPunct{\mcitedefaultmidpunct}
{\mcitedefaultendpunct}{\mcitedefaultseppunct}\relax
\EndOfBibitem
\bibitem[Duboué-Dijon \latin{et~al.}(2018)Duboué-Dijon, Mason, Fischer, and
  Jungwirth]{DDijon2018}
Duboué-Dijon,~E.; Mason,~P.~E.; Fischer,~H.~E.; Jungwirth,~P. Hydration and
  Ion Pairing in Aqueous Mg$^{2+}$ and Zn$^{2+}$ Solutions: Force-Field
  Description Aided by Neutron Scattering Experiments and Ab Initio Molecular
  Dynamics Simulations. \emph{J. Phys. Chem. B} \textbf{2018}, \emph{122},
  3296--3306\relax
\mciteBstWouldAddEndPuncttrue
\mciteSetBstMidEndSepPunct{\mcitedefaultmidpunct}
{\mcitedefaultendpunct}{\mcitedefaultseppunct}\relax
\EndOfBibitem
\bibitem[Duboué-Dijon \latin{et~al.}(2020)Duboué-Dijon, Javanainen, Delcroix,
  Jungwirth, and Martinez-Seara]{DDijon2020}
Duboué-Dijon,~E.; Javanainen,~M.; Delcroix,~P.; Jungwirth,~P.;
  Martinez-Seara,~H. A Practical Guide to Biologically Relevant Molecular
  Simulations with Charge Scaling for Electronic Polarization. \emph{J. Chem.
  Phys.} \textbf{2020}, \emph{153}, 050901\relax
\mciteBstWouldAddEndPuncttrue
\mciteSetBstMidEndSepPunct{\mcitedefaultmidpunct}
{\mcitedefaultendpunct}{\mcitedefaultseppunct}\relax
\EndOfBibitem
\bibitem[Pyykkö(2018)]{Pyykko2017}
Pyykkö,~P. Year-2017 Nuclear Quadrupole Moments. \emph{Mol. Phys.}
  \textbf{2018}, \emph{116}, 1328--1338\relax
\mciteBstWouldAddEndPuncttrue
\mciteSetBstMidEndSepPunct{\mcitedefaultmidpunct}
{\mcitedefaultendpunct}{\mcitedefaultseppunct}\relax
\EndOfBibitem
\bibitem[Hertz \latin{et~al.}(1974)Hertz, Holz, Keller, Versmold, and
  Yoon]{HertzNaLi1974}
Hertz,~H.~G.; Holz,~M.; Keller,~G.; Versmold,~H.; Yoon,~C. Nuclear Magnetic
  Relaxation of Alkali Metal Ions in Aqueous Solutions. \emph{Ber. Bunsenges.
  Phys. Chem.} \textbf{1974}, \emph{78}, 493--509\relax
\mciteBstWouldAddEndPuncttrue
\mciteSetBstMidEndSepPunct{\mcitedefaultmidpunct}
{\mcitedefaultendpunct}{\mcitedefaultseppunct}\relax
\EndOfBibitem
\bibitem[Fumino \latin{et~al.}(1997)Fumino, Shimizu, and Taniguchi]{Fumino1997}
Fumino,~K.; Shimizu,~A.; Taniguchi,~Y. Concentration Dependence of Nuclear
  Magnetic Relaxation Rates of Alkali Metal Ionic Nuclei in Dilute Solutions.
  \emph{Denki Kagaku} \textbf{1997}, \emph{65}, 198--203\relax
\mciteBstWouldAddEndPuncttrue
\mciteSetBstMidEndSepPunct{\mcitedefaultmidpunct}
{\mcitedefaultendpunct}{\mcitedefaultseppunct}\relax
\EndOfBibitem
\bibitem[Weingärtner and Hertz(1977)Weingärtner, and Hertz]{Weingaertner1977}
Weingärtner,~H.; Hertz,~H.~G. Magnetic Relaxation by Quadrupolar Interaction
  of Ionic Nuclei in Non-Aqueous Electrolyte Solutions. Part I. Limiting Values
  for Infinite Dilution. \emph{Ber. Bunsenges. Phys. Chem.} \textbf{1977},
  \emph{81}, 1204--1221\relax
\mciteBstWouldAddEndPuncttrue
\mciteSetBstMidEndSepPunct{\mcitedefaultmidpunct}
{\mcitedefaultendpunct}{\mcitedefaultseppunct}\relax
\EndOfBibitem
\bibitem[Struis \latin{et~al.}(1989)Struis, De~Bleijser, and Leyte]{Struis1989}
Struis,~R. P. W.~J.; De~Bleijser,~J.; Leyte,~J.~C. Magnesium-25(2+) and
  Chloride-35 Quadrupolar Relaxation in Aqueous Magnesium Chloride Solutions at
  25$^{\circ}${C}. 1. {L}imiting Behavior for Infinite Dilution. \emph{J. Phys.
  Chem.} \textbf{1989}, \emph{93}, 7932--7942\relax
\mciteBstWouldAddEndPuncttrue
\mciteSetBstMidEndSepPunct{\mcitedefaultmidpunct}
{\mcitedefaultendpunct}{\mcitedefaultseppunct}\relax
\EndOfBibitem
\bibitem[Helm and Hertz(1981)Helm, and Hertz]{HelmHertz1981}
Helm,~L.; Hertz,~H.~G. The Hydration of the Alkaline Earth Metal Ions
  {Mg}$^{2+}$, {Ca}$^{2+}$, {Sr}$^{2+}$ and {Ba}$^{2+}$, a Nuclear Magnetic
  Relaxation Study Involving the Quadrupole Moment of the Ionic Nuclei.
  \emph{Z. Phys. Chem.} \textbf{1981}, \emph{127}, 23--44\relax
\mciteBstWouldAddEndPuncttrue
\mciteSetBstMidEndSepPunct{\mcitedefaultmidpunct}
{\mcitedefaultendpunct}{\mcitedefaultseppunct}\relax
\EndOfBibitem
\bibitem[Chaker \latin{et~al.}(2019)Chaker, Salanne, Delaye, and
  Charpentier]{Chaker2019}
Chaker,~Z.; Salanne,~M.; Delaye,~J.-M.; Charpentier,~T. {NMR} Shifts in
  Aluminosilicate Glasses via Machine Learning. \emph{Phys. Chem. Chem. Phys.}
  \textbf{2019}, \emph{21}, 21709--21725\relax
\mciteBstWouldAddEndPuncttrue
\mciteSetBstMidEndSepPunct{\mcitedefaultmidpunct}
{\mcitedefaultendpunct}{\mcitedefaultseppunct}\relax
\EndOfBibitem
\bibitem[Paruzzo \latin{et~al.}(2018)Paruzzo, Hofstetter, Musil, De, Ceriotti,
  and Emsley]{Paruzzo2018}
Paruzzo,~F.~M.; Hofstetter,~A.; Musil,~F.; De,~S.; Ceriotti,~M.; Emsley,~L.
  Chemical Shifts in Molecular Solids by Machine Learning. \emph{Nat. Commun.}
  \textbf{2018}, \emph{9}, 4501\relax
\mciteBstWouldAddEndPuncttrue
\mciteSetBstMidEndSepPunct{\mcitedefaultmidpunct}
{\mcitedefaultendpunct}{\mcitedefaultseppunct}\relax
\EndOfBibitem
\bibitem[Kananenka \latin{et~al.}(2019)Kananenka, Yao, Corcelli, and
  Skinner]{Kananenka2019}
Kananenka,~A.~A.; Yao,~K.; Corcelli,~S.~A.; Skinner,~J.~L. Machine Learning for
  Vibrational Spectroscopic Maps. \emph{J. Chem. Theor. Comput.} \textbf{2019},
  \emph{15}, 6850--6858\relax
\mciteBstWouldAddEndPuncttrue
\mciteSetBstMidEndSepPunct{\mcitedefaultmidpunct}
{\mcitedefaultendpunct}{\mcitedefaultseppunct}\relax
\EndOfBibitem
\bibitem[Baiz \latin{et~al.}(2020)Baiz, Błasiak, Bredenbeck, Cho, Choi,
  Corcelli, Dijkstra, Feng, Garrett-Roe, Ge, Hanson-Heine, Hirst, Jansen, Kwac,
  Kubarych, Londergan, Maekawa, Reppert, Saito, Roy, Skinner, Stock, Straub,
  Thielges, Tominaga, Tokmakoff, Torii, Wang, Webb, and Zanni]{Baiz2021}
Baiz,~C.~R.; Błasiak,~B.; Bredenbeck,~J.; Cho,~M.; Choi,~J.-H.;
  Corcelli,~S.~A.; Dijkstra,~A.~G.; Feng,~C.-J.; Garrett-Roe,~S.; Ge,~N.-H.;
  Hanson-Heine,~M. W.~D.; Hirst,~J.~D.; Jansen,~T. L.~C.; Kwac,~K.;
  Kubarych,~K.~J.; Londergan,~C.~H.; Maekawa,~H.; Reppert,~M.; Saito,~S.;
  Roy,~S.; Skinner,~J.~L.; Stock,~G.; Straub,~J.~E.; Thielges,~M.~C.;
  Tominaga,~K.; Tokmakoff,~A.; Torii,~H.; Wang,~L.; Webb,~L.~J.; Zanni,~M.~T.
  Vibrational Spectroscopic Map, Vibrational Spectroscopy, and Intermolecular
  Interaction. \emph{Chem. Rev.} \textbf{2020}, \emph{120}, 7152--7218\relax
\mciteBstWouldAddEndPuncttrue
\mciteSetBstMidEndSepPunct{\mcitedefaultmidpunct}
{\mcitedefaultendpunct}{\mcitedefaultseppunct}\relax
\EndOfBibitem
\end{mcitethebibliography}

\providecommand{\latin}[1]{#1}
\makeatletter
\providecommand{\doi}
  {\begingroup\let\do\@makeother\dospecials
  \catcode`\{=1 \catcode`\}=2 \doi@aux}
\providecommand{\doi@aux}[1]{\endgroup\texttt{#1}}
\makeatother
\providecommand*\mcitethebibliography{\thebibliography}
\csname @ifundefined\endcsname{endmcitethebibliography}
  {\let\endmcitethebibliography\endthebibliography}{}

\clearpage

\begin{figure*}[t!]
\centering
\includegraphics[width=0.99\linewidth]{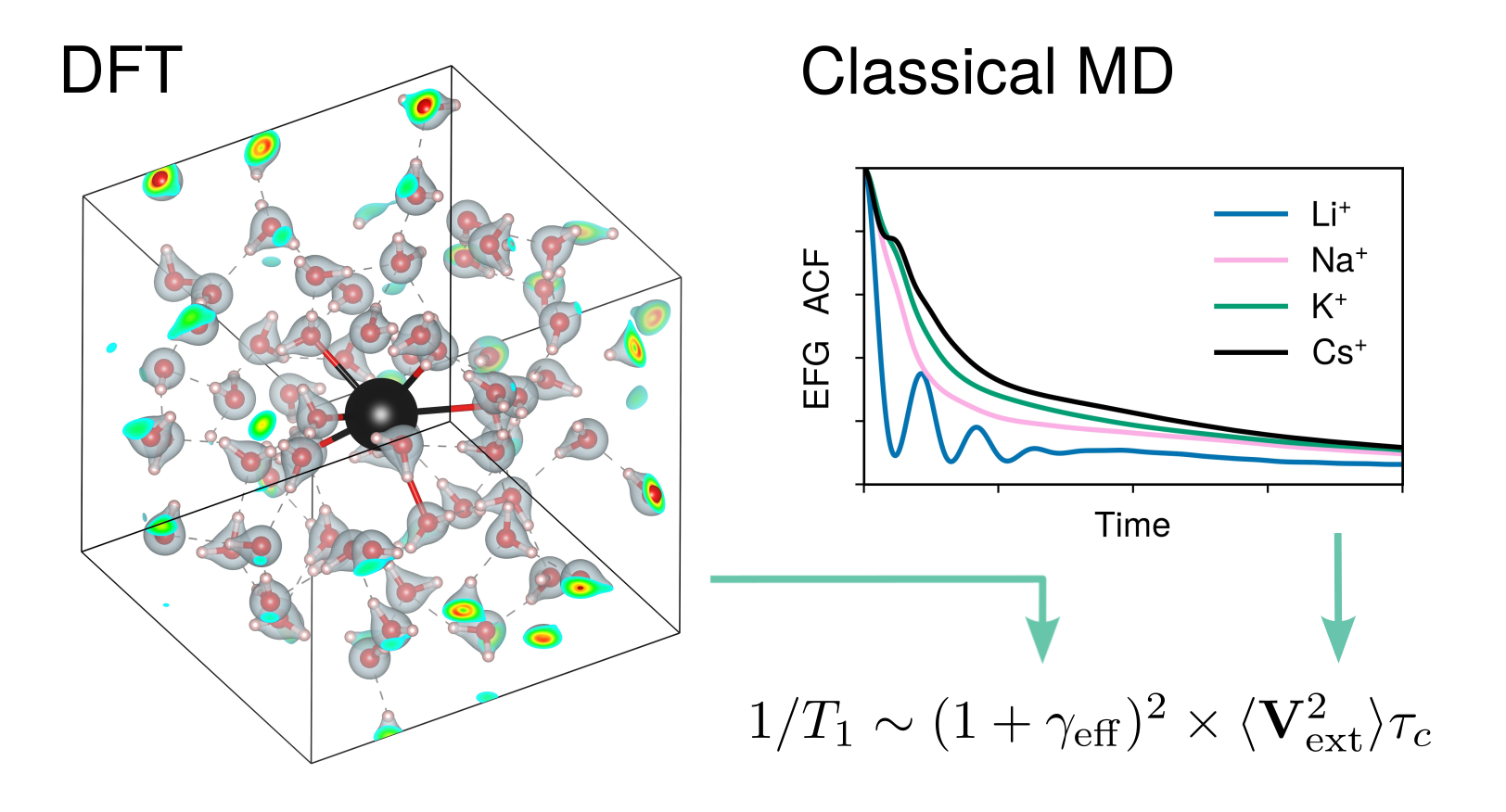}
\caption{For Table of Contents Only.}
\end{figure*}

\clearpage

\renewcommand\thesection{S\arabic{section}}
\renewcommand\thefigure{S\arabic{figure}}
\renewcommand\thetable{S\arabic{table}}   
\setcounter{figure}{0}
\setcounter{table}{0}
\setcounter{section}{0}

\clearpage
\section*{Supporting Information}
\section{Details on the Sternheimer approximation}

In the present work, the system dynamics was generated using classical, force-field based molecular dynamics (FF MD). In principle, the most accurate way of calculating the quadrupolar NMR relaxation rates using FF MD
would be by evaluating the ab initio (AI) electric field gradients (EFGs) ${\mathbf{V}}_{\rm AI}$ at the ion position on every system configuration of interest. The 
dynamics of  ${\mathbf{V}}_{\rm AI}$  fluctuations is 
then given by the variance $\langle {\mathbf{V}}^2_{\rm AI} \rangle$ and the effective correlation time 
$\tau_{\rm c, \,AI}$:
\begin{equation}
\tau_{\rm c, \,AI} =
\langle {\mathbf{V}}^2_{\rm AI} \rangle^{-1} \, 
\int_0^{\infty} {\rm d}t \,
\langle {\mathbf{V}}_{\rm AI}(t)\,{:}{\mathbf{V}}_{\rm AI}(0) \rangle.
\end{equation}
Note that the dynamics of AI EFGs computed on system configurations generated using ab initio molecular dynamics (AIMD) might somewhat differ from that with FF MD. 

The calculation of AI EFGs ${\mathbf{V}}_{\rm AI}$ can be quite expensive 
if done for a large number of configurations. In practice, here we employed the Sternheimer approximation that relates the classical, external EFGs ${\mathbf{V}}_{\rm ext}$ computed with point charges 
and dipoles of a FF at hand to ${\mathbf{V}}_{\rm AI}$
via the effective Sternheimer factor 
$\gamma_{\rm eff}$:
\begin{equation}
\label{eq:SASI}
{\mathbf{V}}_{\rm AI} \simeq (1 + \gamma_{\rm eff}) 
{\mathbf{V}}_{\rm ext}
\end{equation}
The dynamics of  ${\mathbf{V}}_{\rm ext}$ 
fluctuations are characterized by the variance $\langle {\mathbf{V}}^2_{\rm ext} \rangle$ and the effective correlation time $\tau_{\rm c, \,ext}$ (simply 
denoted by $\tau_{\rm c}$ in the main text):
\begin{equation}
\tau_{\rm c, \,ext} =
\langle {\mathbf{V}}^2_{\rm ext} \rangle^{-1} \, 
\int_0^{\infty} {\rm d}t \,
\langle {\mathbf{V}}_{\rm ext}(t)\,{:}{\mathbf{V}}_{\rm ext}(0) \rangle.
\end{equation}

Nevertheless, the Sternheimer approximation does not 
perfectly capture ${\mathbf{V}}_{\rm AI}$ (Fig. 4 of the main text), i.e. there exists an error term $\Delta {\mathbf{V}}$:
\begin{equation}
{\mathbf{V}}_{\rm AI}(t) = (1 + \gamma_{\rm eff}) {\mathbf{V}}_{\rm ext}(t)
+ \Delta {\mathbf{V}}(t)
\end{equation}
The ACF of AI EFG can thus be recast as follows:
\begin{equation}
\label{eq:vai_detailed}
\begin{split}
\langle {\mathbf{V}}_{\rm AI}(t)\,{:}{\mathbf{V}}_{\rm AI}(0) \rangle &= (1 + \gamma_{\rm eff})^2 
\langle {\mathbf{V}}_{\rm ext}(t)\,{:}{\mathbf{V}}_{\rm ext}(0)\rangle + \langle \Delta {\mathbf{V}}(t)\,{:}
\Delta {\mathbf{V}}(0) \rangle \\
&+ 2(1 + \gamma_{\rm eff}) 
\langle {\mathbf{V}}_{\rm ext}(t)\,{:}
\Delta {\mathbf{V}}(0) \rangle.
\end{split}
\end{equation}
Finally, by integrating Eq.~\eqref{eq:vai_detailed}
over time, we find 
\begin{equation}
\label{eq:vai}
\begin{split}
 \langle {\mathbf{V}}^2_{\rm AI} \rangle \, \tau_{\rm c, \, AI} &= (1 + \gamma_{\rm eff})^2 \langle {\mathbf{V}}^2_{\rm ext} \rangle \, \tau_{\rm c, \, ext} + \langle \Delta {\mathbf{V}}^2 \rangle \, \tau_{\rm c, \, \Delta} + 2(1 + \gamma_{\rm eff}) \int_0^{\infty} {\rm d}t \,
\langle {\mathbf{V}}_{\rm ext}(t)\,{:}
\Delta {\mathbf{V}}(0) \rangle,
\end{split}
\end{equation}
where $\langle \Delta {\mathbf{V}}^2 \rangle$ 
is the variance of the error term $\Delta {\mathbf{V}}$
and $\tau_{\rm c, \,\Delta}$ is its 
effective correlation time 
\begin{equation}
\tau_{\rm c, \,\Delta} =
\langle {\Delta \mathbf{V}}^2 \rangle^{-1} \, 
\int_0^{\infty} {\rm d}t \,
\langle {\Delta \mathbf{V}}(t)\,{:}{\Delta \mathbf{V}}(0) \rangle.
\end{equation}
Evidently from Eq.~\eqref{eq:vai}, both the variance of 
$\langle {\mathbf{V}}^2_{\rm AI} \rangle$
and the correlation time of the AI EFG 
$\tau_{\rm c, \,AI}$ are 
affected by the fluctuations of the error term
$\Delta \mathbf{V}$.

\begin{figure*}[p!]
\centering
\includegraphics[width=0.49\linewidth]{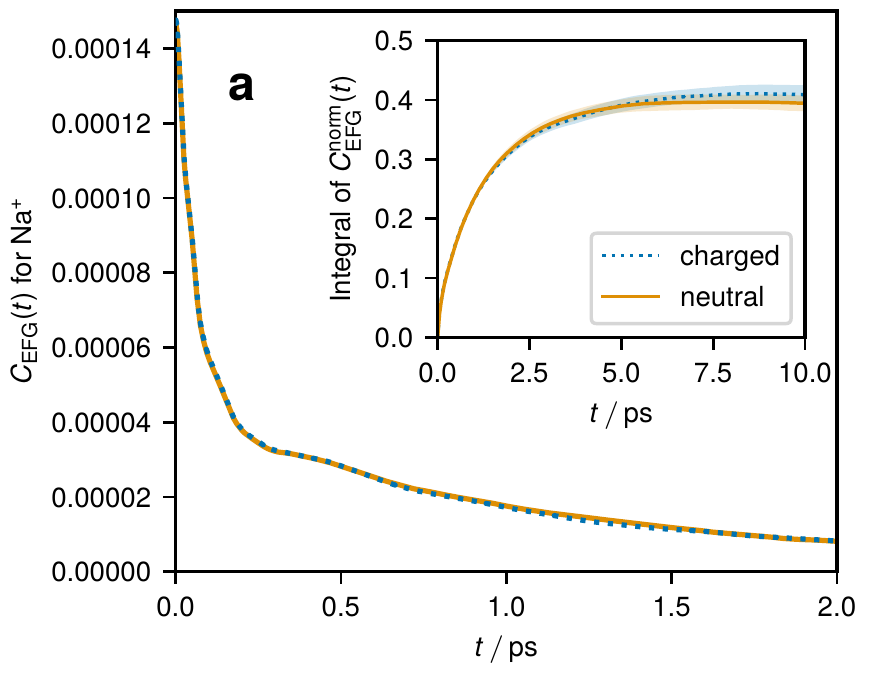}
\includegraphics[width=0.49\linewidth]{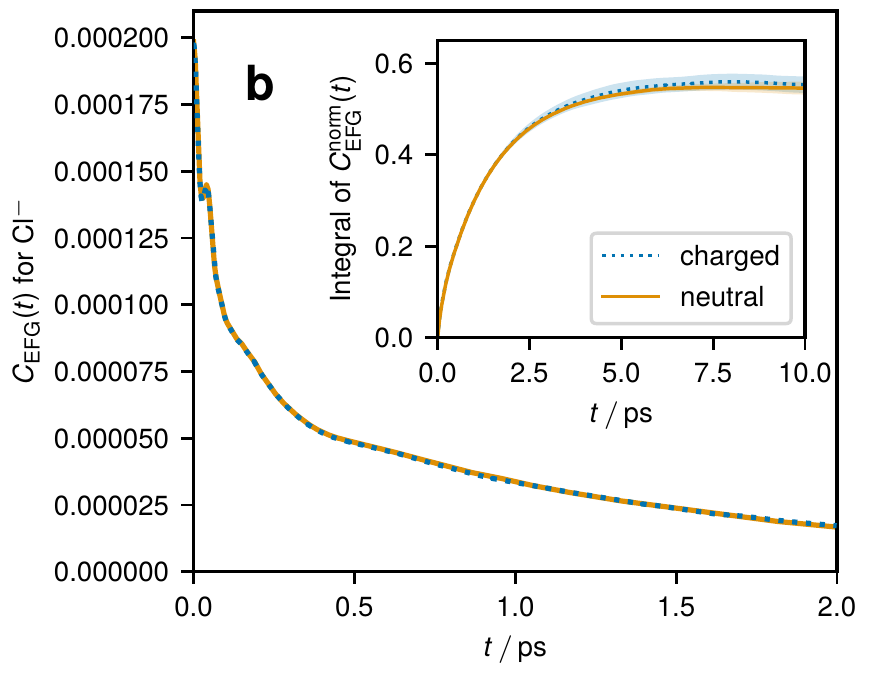}

\vspace*{2mm}

\includegraphics[width=0.49\linewidth]{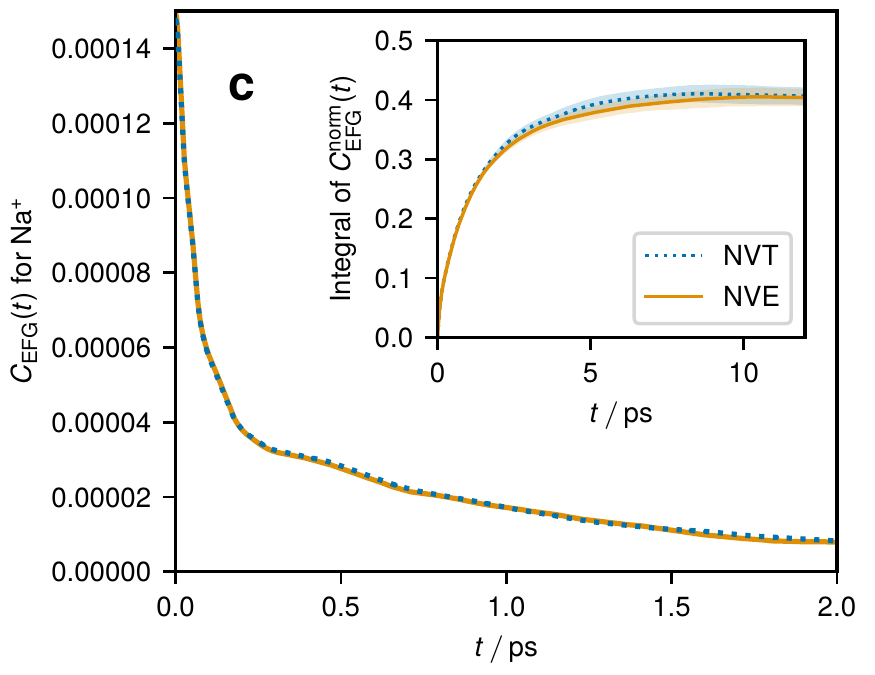}
\includegraphics[width=0.49\linewidth]{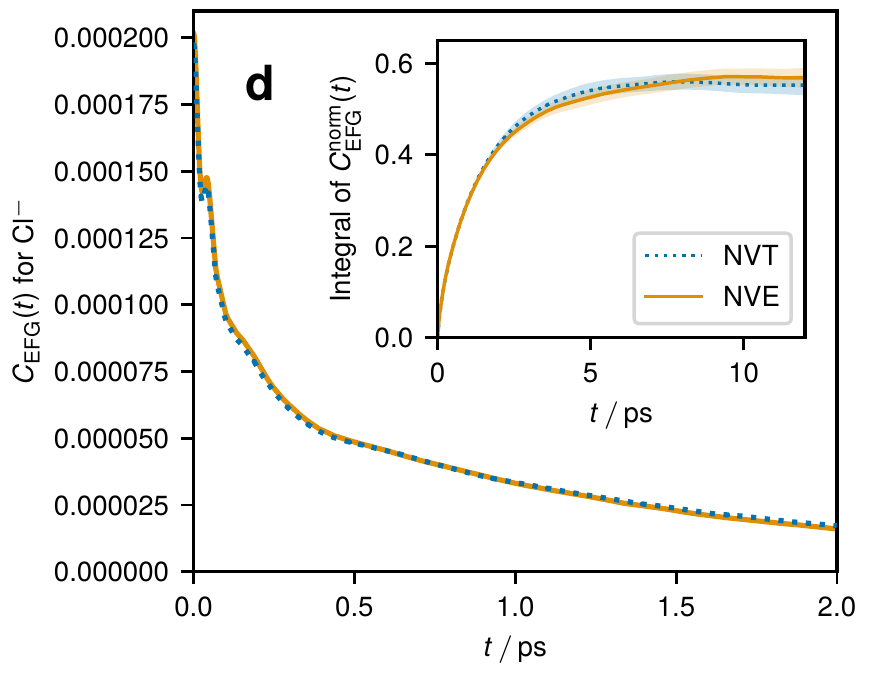}
\caption{Simulation details. 
Comparison between the EFG ACF and the integral of the normalized EFG ACF (inset)
for Na$^+$ ({\bf a}) and Cl$^-$ ({\bf b}) as simulated in a smaller non-electroneutral
system (blue dashed lines)
with $N=256$ water molecules and 1 ion and a larger electroneutral system (solid yellow lines) with $N=1000$ water molecules and 1 NaCl ion pair. For the smaller systems, 5 independent simulation runs of length 1 ns were performed. For the larger system, the results were extracted from 1 run of length 10 ns.
Comparison between the EFG ACF and the integral of the normalized EFG ACF (inset)
for Na$^+$ ({\bf c}) and Cl$^-$ ({\bf d}) as sampled in the $NVT$ (blue dashed lines)
and $NVE$ (solid yellow lines) ensemble. The $NVT$ runs correspond to the smaller system
in {\bf a} and {\bf b}. The $NVE$ runs were performed in a single 5 ns simulation for the 
system with $N=256$ water molecules using an equilibrated system state form $NVT$
simulation as a starting configuration.
In all cases, the ions and water were simulated using the Madrid-2019 model.
}
\end{figure*}

\clearpage

\vspace*{3.5cm}
\begin{figure*}[h!]
\centering
\includegraphics[width=0.99\linewidth]{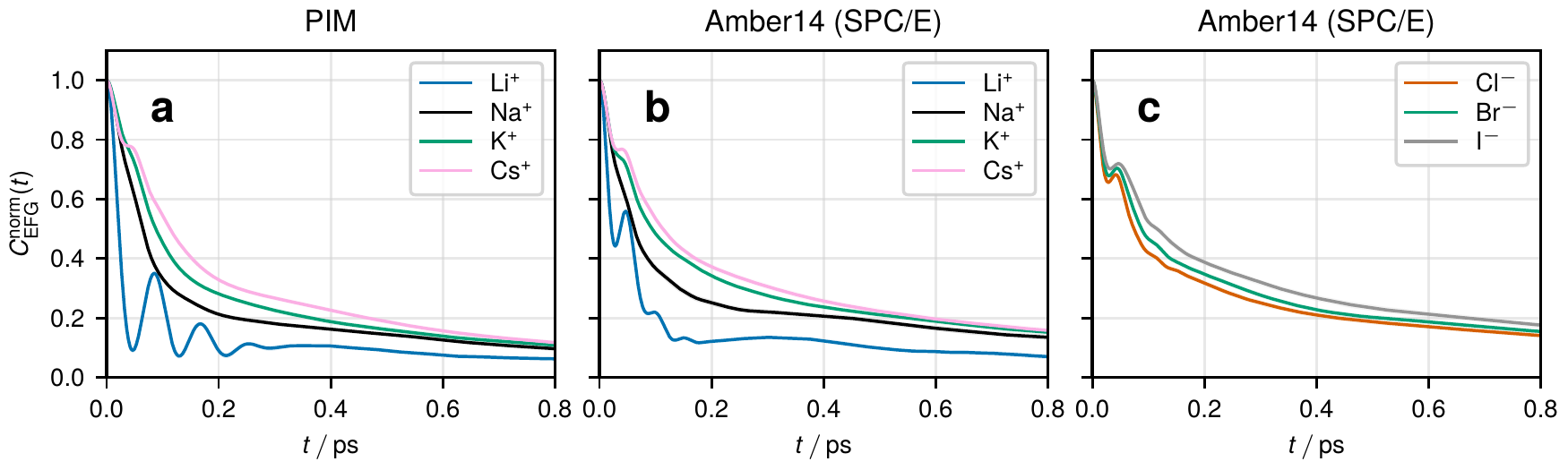}
\includegraphics[width=0.98\linewidth]{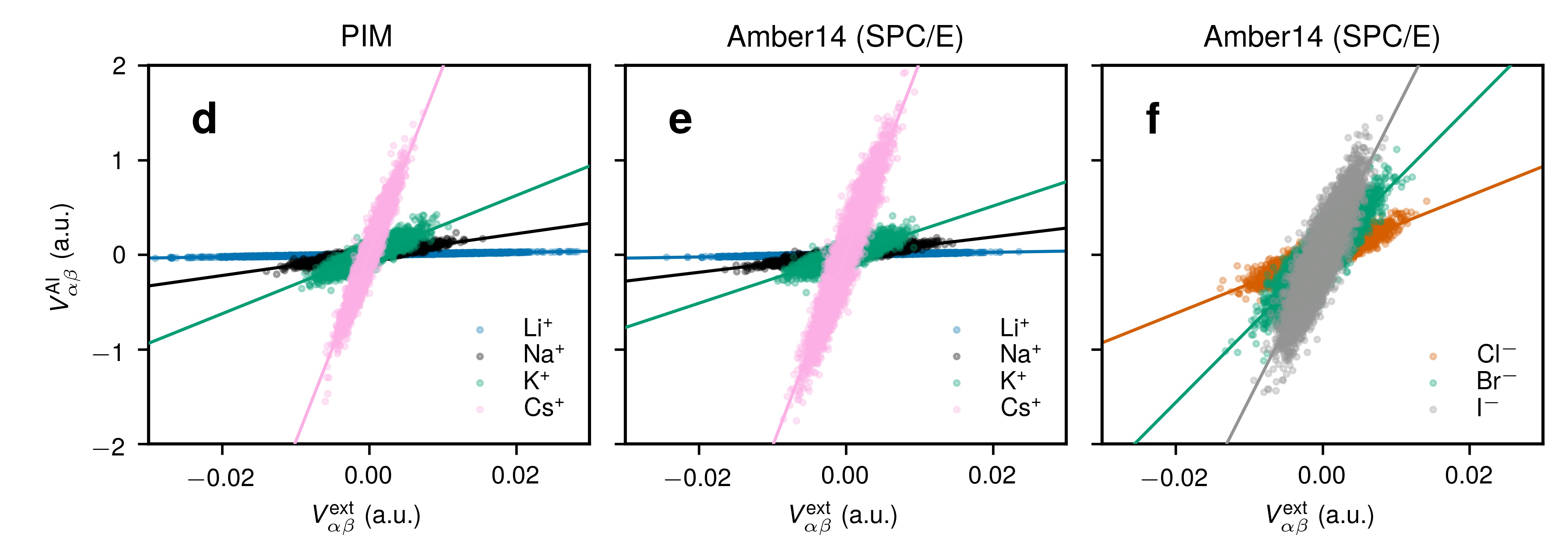}
\caption{Normalized EFG ACFs for alkali metal (({\bf a}) in 
the PIM and ({\bf b}) in the Amber14 FF) and 
halide ions (({\bf c}) in the Amber14 FF).
The corresponding validation of the Sternheimer approximation
for the systems in ({\bf a}), ({\bf b}), and ({\bf c})
is shown in ({\bf d}), ({\bf e}), and ({\bf f}),
respectively.}
\end{figure*}

\clearpage

\begin{figure*}[p!]
\centering
\includegraphics[width=0.9\linewidth]{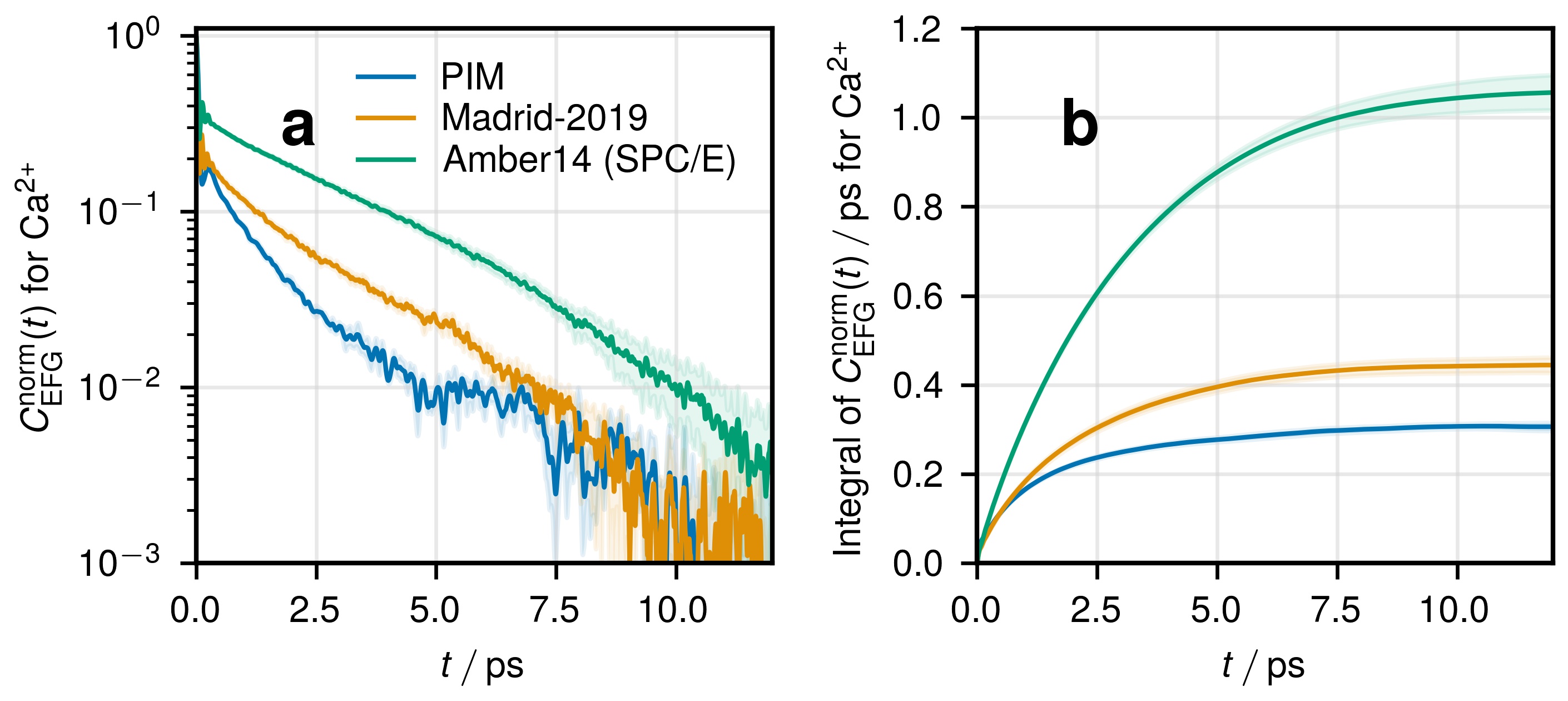}
\caption{Comparison between the normalized EFG ACFs ({\bf a}) 
and their integrals ({\bf b}) for Ca$^{2+}$ in the PIM (blue lines), 
Madrid-2019 model (yellow lines), and  the Amber14 FF (green lines).
({\bf a}) and ({\bf b}) share the same legend shown in ({\bf a}).}
\end{figure*}

\clearpage

\begin{figure*}[p!]
\centering
\includegraphics[width=0.99\textwidth]{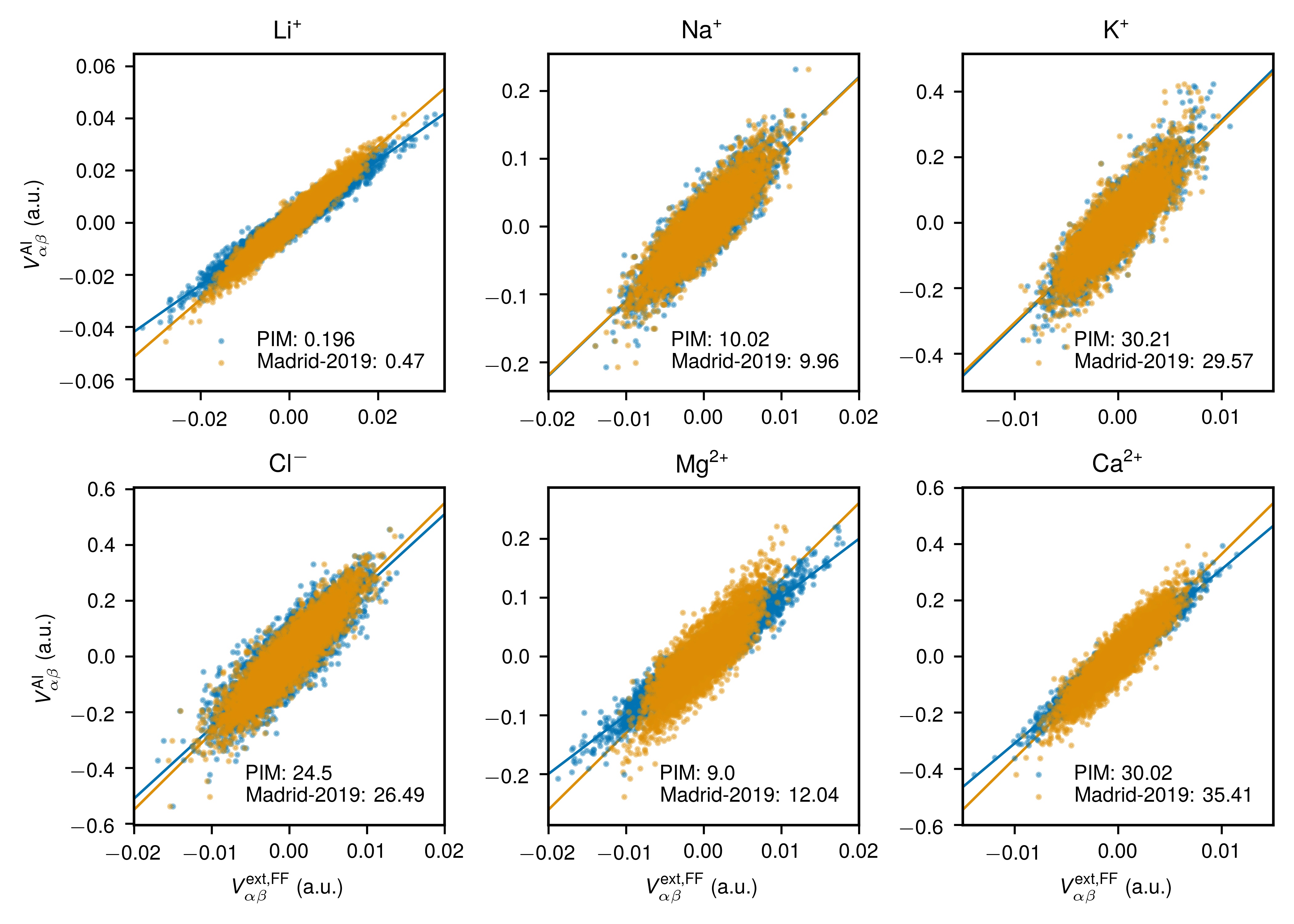}
\caption{Effect of changing the representation of the charge distribution in the classical force field on the 
resulting effective Sternheimer factors in the PIM model for various ions. The AI EFGs
$V_{\alpha\beta}^{\rm AI}$ obtained on the set of configurations generated using PIM dynamics are plotted against the external, force field specific EFG $V_{\alpha\beta}^{\rm ext, FF}$ as computed using the PIM (blue dots) or using the Madrid-2019 FF (yellow dots) on the same set of configurations. The resulting
Sternheimer factors $\gamma_{\rm eff}$, 
$V_{\alpha\beta}^{\rm AI} = (1+\gamma_{\rm eff})  V_{\alpha\beta}^{\rm ext, FF}$ are listed in the legends.
}
\end{figure*}

\clearpage

\begin{figure*}[p!]
\centering
\includegraphics[width=0.99\textwidth]{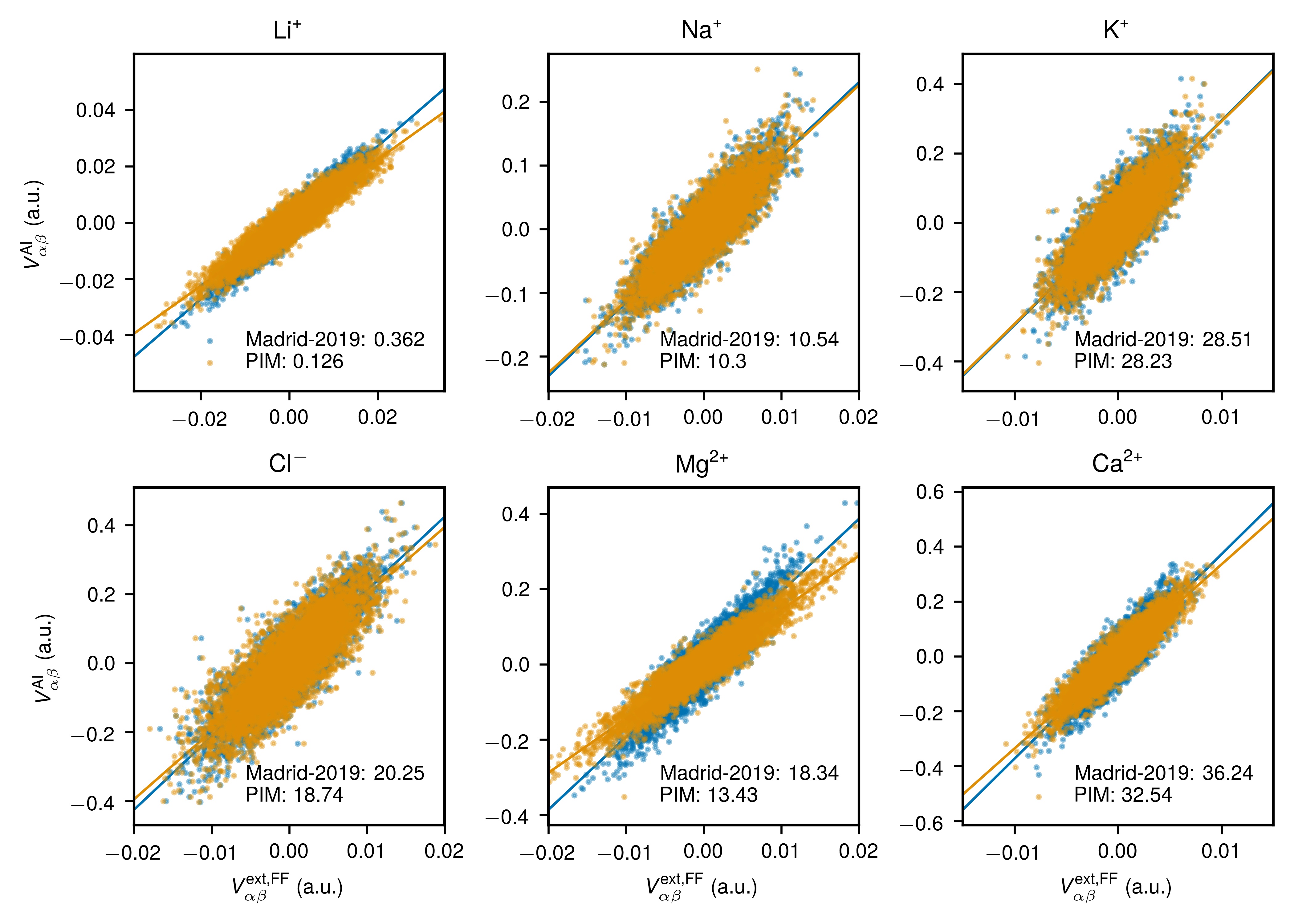}
\caption{Effect of changing the representation of the charge distribution in the classical force field on the 
resulting effective Sternheimer factors in the Madrid-2019 model for various ions. The AI EFGs
$V_{\alpha\beta}^{\rm AI}$ obtained on the set of configurations generated using Madrid-2019 dynamics are plotted against the external, force field specific EFG $V_{\alpha\beta}^{\rm ext, FF}$ as computed using the Madrid-2019 FF (blue dots) or using the PIM (yellow dots) on the same set of configurations. The resulting
Sternheimer factors $\gamma_{\rm eff}$, 
$V_{\alpha\beta}^{\rm AI} = (1+\gamma_{\rm eff}) V_{\alpha\beta}^{\rm ext, FF}$ are listed in the legends.}
\end{figure*}

\clearpage

\begin{figure*}[p!]
\centering
\includegraphics[width=0.99\textwidth]{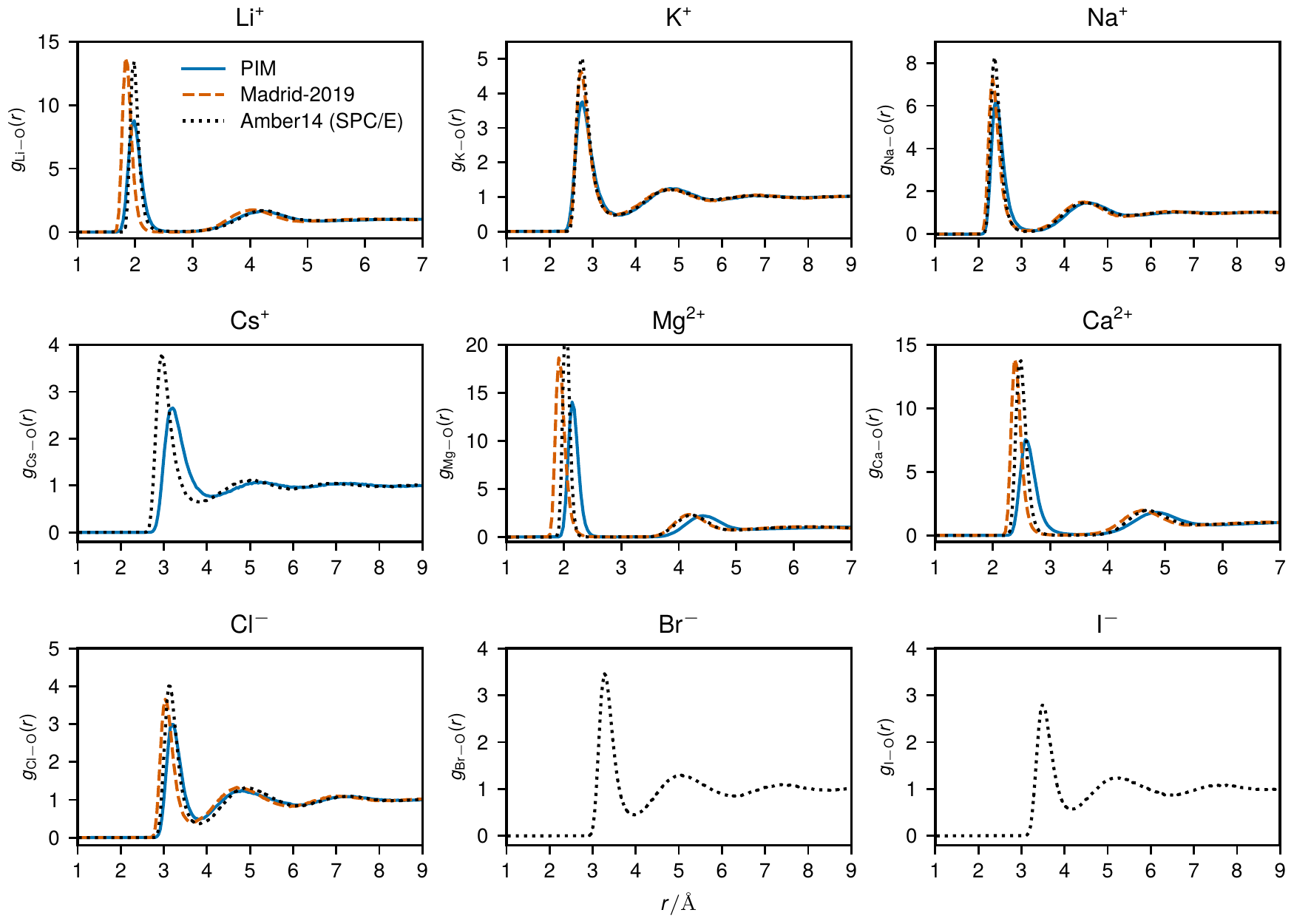}
\caption{Ion-oxygen radial distribution functions in the PIM (solid blue lines), Madrid-2019 (dashed red lines),
and Amber14 FF parameters for ions and SPC/E water (dotted black lines) for different ions at infinite dilution.}
\end{figure*}

\begin{figure*}[p!]
\centering
\includegraphics[width=0.99\textwidth]{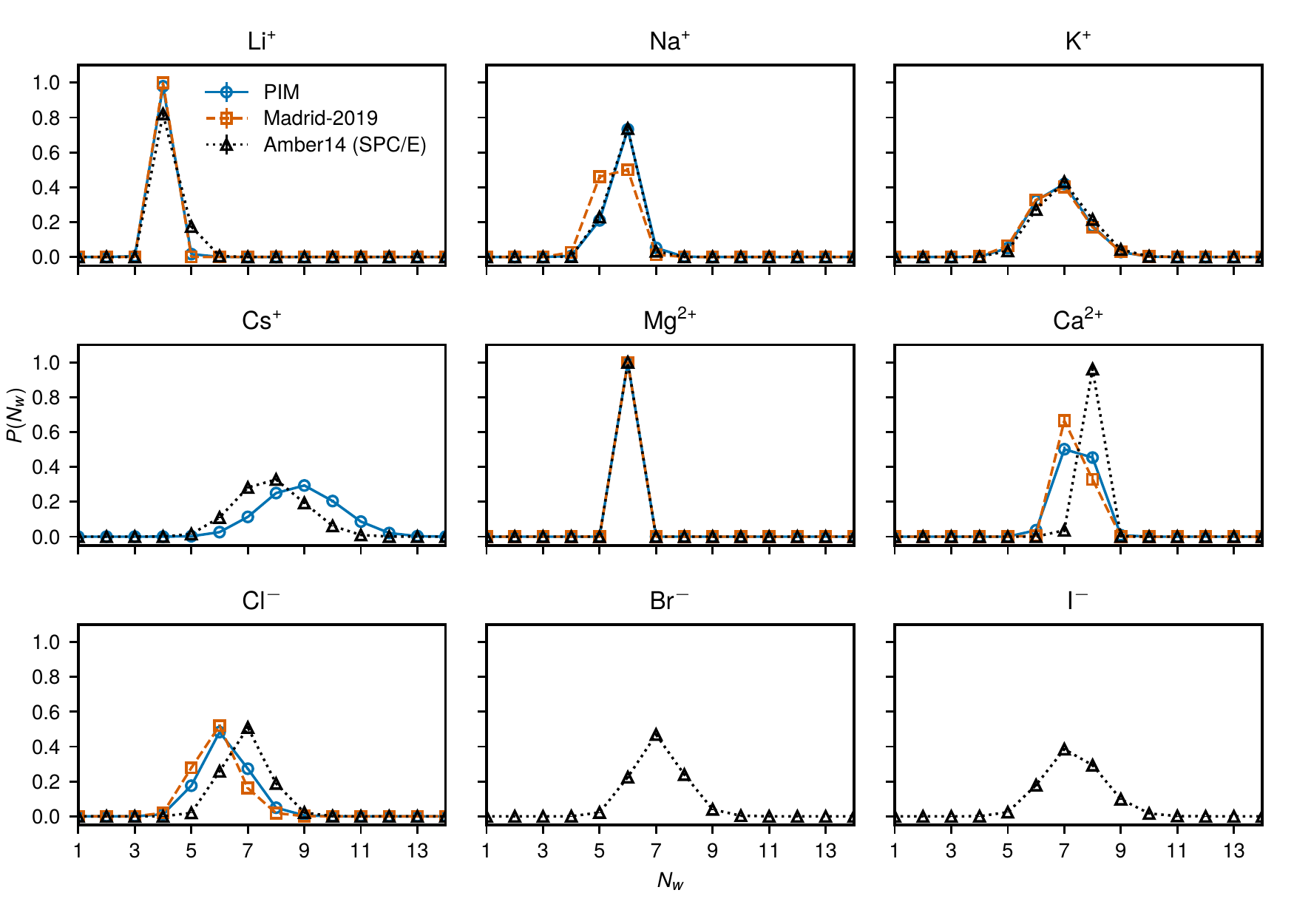}
\caption{Probability of finding $N_{\rm w}$ water molecules in the first hydration shell of single ions in the PIM (blue open circles), the Madrid-2019 model (red open squares), and
Amber14 FF parameters for ions and SPC/E water (black open triangles).}
\end{figure*}

\clearpage

\begin{table*}[p!]
\caption{Structure of the first hydration shell in different classical FFs. The first minimum of $g_{{\rm ion-O}}(r)$,
$r_{1, {\rm min}}$, is used to estimate the first shell coordination number $CN_1 = 4\pi \rho_{\rm O} \int_0^{r_{1, {\rm min}}} {\rm d}r \, r^2 g_{{\rm ion-O}}(r)$ with $\rho_{\rm O}$ the number density of oxygen atoms in the simulation box.}
\vspace*{2mm}
\begin{tabular}{|c|c|c|c|c|c|c|c|}
\hline
Ion & $r^{\rm PIM}_{1, {\rm min}}$ (\AA) & $CN^{\rm PIM}_1$ &
$r^{\rm Madrid-2019}_{1, {\rm min}}$ (\AA) & $CN^{\rm Madrid-2019}_1$ &
$r^{\rm Amber14}_{1, {\rm min}}$ (\AA) & $CN^{\rm Amber14}_1$ \\ \hline
Li$^+$  & 2.68 & 4.0 & 2.69 & 4.0 & 2.67 & 4.2 \\ \hline
Na$^+$  & 3.28 & 5.8 & 3.15 & 5.5 & 3.17 & 5.8 \\ \hline
K$^+$   & 3.63 & 6.8 & 3.53 & 6.8 & 3.53 & 7.0 \\ \hline
Cs$^+$  & 4.12 & 8.9 & --- & --- & 3.82 & 7.8 \\ \hline
Mg$^{2+}$ & 2.96 & 6.0 & 3.00 & 6.0 & 2.88 & 6.0 \\ \hline
Ca$^{2+}$ & 3.53 & 7.4 & 3.24 & 7.3 & 3.22 & 8.0 \\ \hline
Cl$^-$  & 3.82 & 6.2 & 3.68 & 5.9 & 3.80 & 6.9 \\ \hline
Br$^-$  & --- & --- & --- & --- & 3.94 & 7.0 \\ \hline
I$^-$  & --- & --- & --- & --- & 4.15 & 7.3 \\ \hline
\end{tabular}
\end{table*}

\clearpage

\begin{table*}[p!]
\caption{\label{tab:gamma_nw}
Effective Sternheimer factors as a function of the number of water
molecules $N_{\rm w}$ in the first solvation shell as obtained for different classical FFs.
The values in the parentheses indicate the standard error obtained using bootstrapping.}
\vspace*{2mm}
%\small
\begin{tabular}{|c|c|c|c|}
\hline
Ion    & $\gamma^{\rm PIM}_{\rm eff} (N_{\rm w})$ 
 & $\gamma^{\rm Madrid-2019}_{\rm eff} (N_{\rm w})$ 
 & $\gamma^{\rm Amber14 \, (SPC/E)}_{\rm eff} (N_{\rm w})$ \\ \hline
 
Na$^+$ & \begin{tabular}{@{}l@{}}5: 10.19(0.19) \\ 6: 9.94(0.14) \end{tabular} &  \begin{tabular}{@{}c@{}}5: 10.46(0.12) \\ 6: 10.55(0.20) \end{tabular} & \begin{tabular}{@{}c@{}}5: 8.65(0.14) \\ 6: 8.18(0.11) \end{tabular} \\ \hline
K$^+$  & \begin{tabular}{@{}c@{}}5: 32.60(0.91) \\ 6: 30.20(0.48) 
\\ 7: 30.67(0.56) \\ 8: 27.31(0.79)  
\end{tabular} & \begin{tabular}{@{}c@{}}5: 30.44(1.00) \\ 6: 29.98(0.50) \\ 7: 27.21(0.54) \\ 8: 25.75(0.98) \end{tabular} & \begin{tabular}{@{}c@{}}5: 26.88(0.85) \\ 6: 25.73(0.49) \\ 7: 24.41(0.45) \\ 8: 23.30(0.70) \end{tabular}\\ \hline
Cs$^+$  & \begin{tabular}{@{}c@{}}7: 199.0(2.0) \\ 8: 198.6(2.0) 
\\ 9: 195.9(2.2) \\ 10: 191.7(3.3)  
\end{tabular} & --- & 
\begin{tabular}{@{}c@{}}6: 205.2(1.9) \\ 7: 201.5(1.8) 
\\ 8: 201.7(2.2) \\ 9: 188.6(3.7)  
\end{tabular}
\\ \hline
Cl$^-$ & \begin{tabular}{@{}c@{}}5: 23.75(0.50) \\ 6: 23.92(0.37) \\ 7: 26.18(0.48) \end{tabular} & \begin{tabular}{@{}c@{}}5: 19.91(0.38) \\ 6: 20.69(0.30) \\ 7: 20.09(0.64) \end{tabular} & 
\begin{tabular}{@{}c@{}}6: 29.69(0.43) \\ 7: 30.03(0.40) \\ 8: 30.11(0.69) \end{tabular} \\ \hline
Br$^-$ & --- & --- & 
\begin{tabular}{@{}c@{}}6: 76.6(1.2) \\ 7: 77.1(1.0) \\ 8:
78.3(1.8) \end{tabular} \\ \hline
I$^-$ & --- & --- & 
\begin{tabular}{@{}c@{}}6: 154.5(2.4) \\ 7: 149.4(2.2) \\ 8:
152.5(3.0) \end{tabular} \\ \hline
Ca$^{2+}$ & \begin{tabular}{@{}c@{}}7: 30.02(0.16) \\ 8: 30.07(0.29) \end{tabular}  & \begin{tabular}{@{}c@{}}7: 36.18(0.33) \\ 8: 35.77(0.73) \end{tabular} & \begin{tabular}{@{}c@{}}7: 30.04(0.49) \\ 8: 29.64(0.26) \end{tabular} \\ \hline
\end{tabular}
\end{table*}

\clearpage

\begin{table*}[h!]
\caption{Parameters of the EFG relaxation for different ions in the PIM, Madrid-2019 model, and  Amber14 FF parameters
for the SPC/E water and ions.
$\langle {\mathbf V}_{\rm ext}^2 \rangle$ is the variance of the 
external EFG at the ion position obtained from classical MD simulations.
$\langle {\mathbf V}^2 \rangle$ is the variance of the full EFG at the ion position obtained using the Sternheimer approximation:
$\langle {\mathbf V}_{\rm SA}^2 \rangle = (1+\gamma_{\rm eff})^2 
\langle {\mathbf V}_{\rm ext}^2 \rangle$. 
$\langle {\mathbf V}^2_{\rm AI} \rangle$ is the variance of the full EFG obtained directly from DFT GIPAW calculations 
on a set of 1000 configurations.
$\tau_{\rm c}$ is the correlation time of the external EFG ACF. The values in the parentheses indicate the standard error over 5 independent simulation runs.}
\vspace*{2mm}
\begin{tabular}{|c|c|c|c|c|}
\hline
Ion & 
$\langle {\mathbf V}_{\rm ext}^2 \rangle$ (a.u.) &
$\langle {\mathbf V}_{\rm SA}^2 \rangle$ (a.u.) & 
$\langle {\mathbf V}_{\rm AI}^2 \rangle$ (a.u.) & 
$\tau_{\rm c}$ (ps)  \\ \hline
 \multicolumn{5}{c}{PIM} \\ \cline{1-5}
Li$^+$  & 5.50(0.01) $10^{-4}$  & 7.86(0.05) $10^{-4}$ & 
8.45(0.20) $10^{-4}$ & 0.20(0.01)  \\ \hline
Na$^+$ & 1.15(0.01) $10^{-4}$ & 1.40(0.03) $10^{-2}$  & 
1.83(0.05) $10^{-2}$ & 0.26(0.01)   \\ \hline
K$^+$  & 6.31(0.01) $10^{-5}$  & 6.14(0.13) $10^{-2}$ &
7.94(0.21) $10^{-2}$ & 0.33(0.01)  \\ \hline
Cs$^+$ & 2.28(0.01) $10^{-5}$ & 8.91(0.10) $10^{-1}$ & 
9.73(0.22) $10^{-1}$ & 0.34(0.01)     \\ \hline
Cl$^-$ & 1.33(0.01) $10^{-4}$ & 8.68(0.17) $10^{-2}$ &
1.21(0.02) $10^{-1}$ & 0.51(0.02)     \\ \hline
Mg$^{2+}$ & 1.83(0.01) $10^{-4}$ & 1.83(0.02) $10^{-2}$ & 
2.12(0.06) $10^{-2}$ & 0.091(0.005)  \\ \hline
Ca$^{2+}$ & 7.19(0.02) $10^{-5}$ & 6.91(0.06) $10^{-2}$ &
7.78(0.17) $10^{-2}$ & 0.31(0.01)  \\ \hline

\multicolumn{5}{c}{Madrid-2019} \\ \cline{1-5}
Li$^+$ & 3.75(0.01) $10^{-4}$ & 6.95(0.06) $10^{-4}$ &
7.58(0.17) $10^{-4}$ & 0.107(0.003) \\ \hline
Na$^+$ & 1.48(0.01) $10^{-4}$ & 1.97(0.04) $10^{-2}$  &
2.59(0.06) $10^{-2}$ & 0.41(0.01)  \\ \hline
K$^+$  & 5.99(0.01) $10^{-5}$  & 5.21(0.12) $10^{-2}$ &
7.58(0.17) $10^{-2}$ & 0.36(0.01)  \\ \hline
Cl$^-$ & 1.98(0.01) $10^{-4}$ & 8.92(0.19) $10^{-2}$  &
1.19(0.02) $10^{-1}$ & 0.56(0.02)  \\ \hline
Mg$^{2+}$ & 1.31(0.01) $10^{-4}$ & 4.89(0.09) $10^{-2}$ & 
6.09(0.18) $10^{-2}$ & 0.100(0.003)  \\ \hline
Ca$^{2+}$ & 5.13(0.04) $10^{-5}$ & 7.11(0.13) $10^{-2}$ & 
8.43(0.18) $10^{-2}$ & 0.44(0.01)  \\ \hline

\multicolumn{5}{c}{Amber14 (SPC/E)} \\ \cline{1-5}
Li$^+$ & 3.30(0.01) $10^{-4}$ & 5.05(0.06) $10^{-4}$ &
5.93(0.12) $10^{-4}$ & 0.21(0.01) \\ \hline
Na$^+$ & 1.39(0.01) $10^{-4}$ & 1.22(0.02) $10^{-2}$ & 
1.68(0.04) $10^{-2}$ & 0.41(0.01) \\ \hline
K$^+$  & 6.71(0.01) $10^{-5}$ & 4.22(0.10) $10^{-2}$ &
6.34(0.15) $10^{-2}$ & 0.49(0.02)     \\ \hline
Cs$^+$ & 4.28(0.01) $10^{-5}$ & 1.76(0.02) & 
2.05(0.04) & 0.50(0.02)     \\ \hline
Cl$^-$ & 1.20(0.01) $10^{-4}$ & 1.15(0.02) $10^{-1}$ & 
1.48(0.03) $10^{-1}$ & 0.40(0.01)  \\ \hline
Br$^-$ & 8.17(0.03) $10^{-5}$ & 4.99(0.01) $10^{-1}$ & 
6.48(0.13) $10^{-1}$ & 0.43(0.01)     \\ \hline
I$^-$ & 4.87(0.04) $10^{-5}$ & 1.14(0.02) & 
1.48(0.03) & 0.49(0.01)     \\ \hline
Mg$^{2+}$ & 8.82(0.01) $10^{-5}$ & 1.25(0.03) $10^{-2}$ & 
1.79(0.05) $10^{-2}$ & 0.066(0.003) \\ \hline
Ca$^{2+}$ & 6.04(0.01) $10^{-5}$ & 5.71(0.09) $10^{-2}$ & 
7.06(0.13) $10^{-2}$ & 1.06(0.04)  \\ \hline
\end{tabular}
\end{table*}

\end{document}